\documentclass[11pt]{article}
\usepackage{amssymb,amsmath,amsfonts}
\usepackage{graphicx}
\usepackage{graphics}
\usepackage{eepic,epsfig}

\@addtoreset{equation}{section}

\makeatother

\begin{document}

\title{Fermionic current densities induced by magnetic flux \\
in a conical space with a circular boundary}
\author{E. R. Bezerra de Mello$^{1}$\thanks{%
E-mail: emello@fisica.ufpb.br},\thinspace\ V. B. Bezerra$^{1}$\thanks{%
E-mail: valdir@fisica.ufpb.br}, \thinspace\ A. A. Saharian$^{1,2}$\thanks{%
E-mail: saharian@ysu.am}, \thinspace\ V. M. Bardeghyan$^{2}$ \\
%EndAName
\\
\textit{$^{1}$Departamento de F\'{\i}sica, Universidade Federal da Para\'{\i}%
ba}\\
\textit{58.059-970, Caixa Postal 5.008, Jo\~{a}o Pessoa, PB, Brazil}\vspace{%
0.3cm}\\
\textit{$^2$Department of Physics, Yerevan State University,}\\
\textit{Alex Manoogian Street, 0025 Yerevan, Armenia}}
\maketitle

\begin{abstract}
We investigate the vacuum expectation value of the fermionic current induced
by a magnetic flux in a (2+1)-dimensional conical spacetime in the presence
of a circular boundary. On the boundary the fermionic field obeys MIT bag
boundary condition. For irregular modes, a special case of boundary
conditions at the cone apex is considered, when the MIT bag boundary
condition is imposed at a finite radius, which is then taken to zero. We
observe that the vacuum expectation values for both charge density and
azimuthal current are periodic functions of the magnetic flux with the
period equal to the flux quantum whereas the expectation value of the radial
component vanishes. For both exterior and interior regions, the expectation
values of the current are decomposed into boundary-free and boundary-induced
parts. For a massless field the boundary-free part in the vacuum expectation
value of the charge density vanishes, whereas the presence of the boundary
induces nonzero charge density. Two integral representations are given for
the boundary-free part in the case of a massive fermionic field for
arbitrary values of the opening angle of the cone and magnetic flux. The
behavior of the induced fermionic current is investigated in various
asymptotic regions of the parameters. At distances from the boundary larger
than the Compton wavelength of the fermion particle, the vacuum expectation
values decay exponentially with the decay rate depending on the opening
angle of the cone. We make a comparison with the results already known from
the literature for some particular cases.
\end{abstract}

\bigskip


\bigskip

\section{Introduction}

Topological defects are inevitably produced during symmetry breaking phase
transitions and play an important role in many fields of physics. They
appear in different condensed matter systems including superfluids,
superconductors and liquid crystals. Moreover, symmetry breaking phase
transitions have several cosmological consequences and, within the framework
of grand unified theories, various types of topological defects are
predicted to be formed in the early universe \cite{Vile85}. They provide an
important link between particle physics and cosmology. Among various types
of topological defects, the cosmic strings are of special interest. They are
candidates to produce a number of interesting physical effects, such as the
generation of gravitational waves, gamma ray bursts and high-energy cosmic
rays. Recently, cosmic strings attract a renewed interest partly because a
variant of their formation mechanism is proposed in the framework of brane
inflation \cite{Sara02}.

In the simplest theoretical model describing the infinite straight cosmic
string the spacetime is locally flat except on the string where it has a
Dirac-delta shaped Riemann curvature tensor. From the point of view of
quantum field theory, the corresponding non-trivial topology induces
non-zero vacuum expectation values for several physical observables.
Explicit calculations for the geometry of a single idealized cosmic string
have been developed for different fields \cite{Hell86}-\cite{Beze10}.
Moreover, vacuum polarization effects by higher-dimensional composite
topological defects constituted by a cosmic string and global monopole are
investigated in Refs. \cite{Beze06Comp} for scalar and fermionic fields. The
geometry of a cosmic string in background of de Sitter spacetime has been
recently considered in \cite{Beze09dS}.

Another type of vacuum polarization arises in the presence of boundaries.
The imposed boundary conditions on quantum fields alter the zero-point
fluctuations spectrum and result in additional shifts in the vacuum
expectation values of physical quantities. This is the well-known Casimir
effect (for a review see \cite{Most97}). Note that the Casimir forces
between material boundaries are presently attracting much experimental
attention \cite{Klim09}. In Refs. \cite{Brev95}-\cite{Beze08Ferm}, both
types of sources for the polarization of the vacuum were studied in the
cases of scalar, electromagnetic and fermionic fields, namely, a cylindrical
boundary and a cosmic string, assuming that the boundary is coaxial with the
string. The case of a scalar field was considered in an arbitrary number of
spacetime dimensions, whereas the problems for the electromagnetic and
fermionic fields were studied in four dimensional spacetime. Continuing in
this line of investigation, in the present paper we study the fermionic
current induced by a magnetic flux in a (2+1)-dimensional conical space with
a circular boundary.

As it is well known, field theoretical models in 2+1 dimensions exhibit a
number of interesting effects, such as parity violation, flavour symmetry
breaking, fractionalization of quantum numbers (see Refs. \cite{Dese82}-\cite%
{Dunn99}). An important aspect is the possibility of giving a topological
mass to the gauge bosons without breaking gauge invariance. Field theories
in 2+1 dimensions provide simple models in particle physics and related
theories also rise in the long-wavelength description of certain planar
condensed matter systems, including models of high-temperature
superconductivity. An interesting application of Dirac theory in 2+1
dimensions recently appeared in nanophysics. In a sheet of hexagons from the
graphite structure, known as graphene, the long-wavelength description of
the electronic states can be formulated in terms of the Dirac-like theory of
massless spinors in (2+1)-dimensional spacetime with the Fermi velocity
playing the role of speed of light (for a review see Ref. \cite{Cast09}).
One-loop quantum effects induced by non-trivial topology of graphene made
cylindrical and toroidal nanotubes have been recently considered in Refs.
\cite{Bell09}. The vacuum polarization in graphene with a topological defect
is investigated in Ref. \cite{Site08} within the framework of
long-wavelength continuum model.

The interaction of a magnetic flux tube with a fermionic field gives rise to
a number of interesting phenomena, such as the Aharonov-Bohm effect, parity
anomalies, formation of a condensate and generation of exotic quantum
numbers. For background Minkowski spacetime, the combined effects of the
magnetic flux and boundaries on the vacuum energy have been studied in Refs.
\cite{Lese98,Bene00}. In the present paper we investigate the vacuum
expectation value of the fermionic current induced by vortex configuration
of a gauge field in a (2+1)-dimensional conical space with a circular
boundary. We assume that on the boundary the fermionic field obeys MIT bag
boundary condition. The induced fermionic current is among the most
important quantities that characterize the properties of the quantum vacuum.
Although the corresponding operator is local, due to the global nature of
the vacuum, this quantity carries an important information about the global
properties of the background spacetime. In addition to describing the
physical structure of the quantum field at a given point, the current acts
as the source in the Maxwell equations. It therefore plays an important role
in modelling a self-consistent dynamics involving the electromagnetic field.

From the point of view of the physics in the region outside the conical
defect core, the geometry considered in the present paper can be viewed as a
simplified model for the non-trivial core. This model presents a framework
in which the influence of the finite core effects on physical processes in
the vicinity of the conical defect can be investigated. In particular, it
enables to specify conditions under which the idealized model with the core
of zero thickness can be used. The corresponding results may shed light upon
features of finite core effects in more realistic models, including those
used for defects in crystals and superfluid helium. In addition, the problem
considered here is of interest as an example with combined topological and
boundary induced quantum effects, in which the vacuum characteristics can be
found in closed analytic form.

The organization of the paper is as follows. In the next section we consider
the complete set of solutions to the Dirac equation in the region outside a
circular boundary on which the field obeys MIT bag boundary condition.
Shrinking the radius of the circle to zero we clarify the structure of the
eigenspinors for the boundary-free geometry. These eigenspinors are used in
Sect. \ref{sec:BoundFree} for the evaluation of the vacuum expectation value
of the fermionic current density. Two integral representations are provided
for the charge density and azimuthal component. In Sect. \ref{sec:ExtFC}, we
consider the vacuum expectation values in the region outside a circular
boundary. They are decomposed into boundary-free and boundary-induced parts.
Rapidly convergent integral representations for the latter are obtained.
Similar investigation for the region inside a circular boundary is presented
in Sect. \ref{sec:Int}. The main results are summarized in Sect. \ref%
{sec:Conc}. In Appendix \ref{sec:IntRep} we derive two integral
representations for the series involving the modified Bessel functions.
These representations are used to obtain the fermionic current densities in
the boundary-free geometry. In Appendix \ref{sec:App2New}, we compare the
results of the present paper, in the special case of a magnetic flux in
(2+1)-dimensional Minkowski spacetime, with those from the literature. In
Appendix \ref{sec:App2} we show that the special mode does not contribute to
the vacuum expectation value of the fermionic current in the region inside a
circular boundary.

\section{Model and the eigenspinors in the exterior region}

\label{sec:Ext}

In this paper we consider a two-component spinor field $\psi $, propagating
on a $(2+1)$-dimensional background spacetime with a conical singularity
described by the line-element%
\begin{equation}
ds^{2}=g_{\mu \nu }dx^{\mu }dx^{\nu }=dt^{2}-dr^{2}-r^{2}d\phi ^{2},
\label{ds21}
\end{equation}%
where $r\geqslant 0$, $0\leqslant \phi \leqslant \phi _{0}$, and the points $%
(r,\phi )$ and $(r,\phi +\phi _{0})$ are to be identified. We are interested
in the change of the vacuum expectation value (VEV) of the fermionic current
induced by a magnetic flux in the presence of a circular boundary concentric
with the apex of the cone.

The dynamics of a massive spinor field is governed by the Dirac equation
\begin{equation}
i\gamma ^{\mu }(\nabla _{\mu }+ieA_{\mu })\psi -m\psi =0\ ,\;\nabla _{\mu
}=\partial _{\mu }+\Gamma _{\mu },  \label{Direq}
\end{equation}%
where $A_{\mu }$ is the vector potential for the external electromagnetic
field. In Eq. (\ref{Direq}), $\gamma ^{\mu }=e_{(a)}^{\mu }\gamma ^{(a)}$
are the $2\times 2$ Dirac matrices in polar coordinates and $\Gamma _{\mu }$
is the spin connection. The latter is defined in terms of the flat space
Dirac matrices, $\gamma ^{(a)}$, by the relation
\begin{equation}
\Gamma _{\mu }=\frac{1}{4}\gamma ^{(a)}\gamma ^{(b)}e_{(a)}^{\nu }e_{(b)\nu
;\mu }\ ,  \label{Gammamu}
\end{equation}%
where $;$ means the standard covariant derivative for vector fields. In the
equations above, $e_{(a)}^{\mu }$, $a=0,1,2$, is the basis tetrad satisfying
the relation $e_{(a)}^{\mu }e_{(b)}^{\nu }\eta ^{ab}=g^{\mu \nu }$, with $%
\eta ^{ab}$ being the Minkowski spacetime metric tensor. We assume that the
field obeys the MIT bag boundary condition on the circle with radius $a$:
\begin{equation}
\left( 1+in_{\mu }\gamma ^{\mu }\right) \psi \big|_{r=a}=0\ ,  \label{BCMIT}
\end{equation}%
where $n_{\mu }$ is the outward oriented normal (with respect to the region
under consideration) to the boundary. In particular, from Eq. (\ref{BCMIT})
it follows that the normal component of the fermion current vanishes at the
boundary, $n_{\mu }\bar{\psi}\gamma ^{\mu }\psi =0$, with $\bar{\psi}=\psi
^{\dagger }\gamma ^{0}$ being the Dirac adjoint and the dagger denotes
Hermitian conjugation. In this section we consider the region $r>a$ for
which $n_{\mu }=-\delta _{\mu }^{1}$.

In (2+1)-dimensional spacetime there are two inequivalent irreducible
representations of the Clifford algebra. In the first one we may choose the
flat space Dirac matrices in the form%
\begin{equation}
\gamma ^{(0)}=\sigma _{3},\;\gamma ^{(1)}=i\sigma _{1},\;\gamma
^{(2)}=i\sigma _{2},  \label{DirMat}
\end{equation}%
with $\sigma _{l}$ being Pauli matrices. In the second representation the
gamma matrices can be taken as $\gamma ^{(0)}=-\sigma _{3}$, $\gamma
^{(1)}=-i\sigma _{1}$, $\gamma ^{(2)}=-i\sigma _{2}$. In what follows we use
the representation (\ref{DirMat}). The corresponding results for the second
representation are obtained by changing the sign of the mass, $m\rightarrow
-m$. For the basis tetrads we use the representation below:%
\begin{eqnarray}
e_{(0)}^{\mu } &=&\left( 1,0,0\right) ,  \notag \\
e_{(1)}^{\mu } &=&\left( 0,\cos (q\phi ),-\sin (q\phi )/r\right) ,
\label{Tetrad} \\
e_{(2)}^{\mu } &=&\left( 0,\sin (q\phi ),\cos (q\phi )/r\right) ,  \notag
\end{eqnarray}%
where the parameter $q$ is related to the opening angle of the cone by the
relation
\begin{equation}
q=2\pi /\phi _{0}.  \label{qu}
\end{equation}%
With this choice, for the Dirac matrices in the coordinate system given by
the line element (\ref{ds21}), we have the following representation
\begin{equation}
\gamma ^{0}=\left(
\begin{array}{cc}
1 & 0 \\
0 & -1%
\end{array}%
\right) ,\;\gamma ^{1}=i\left(
\begin{array}{cc}
0 & e^{-iq\phi } \\
e^{iq\phi } & 0%
\end{array}%
\right) ,\;\gamma ^{2}=\frac{1}{r}\left(
\begin{array}{cc}
0 & e^{-iq\phi } \\
-e^{iq\phi } & 0%
\end{array}%
\right) .  \label{DirMat2}
\end{equation}%
Consequently, the Dirac equation takes the form%
\begin{equation}
\Big[\gamma ^{\mu }\left( \partial _{\mu }+ieA_{\mu }\right) +\frac{1-q}{2r}%
\gamma ^{1}+im\Big]\psi =0,  \label{Direq3}
\end{equation}%
where the term with $\gamma ^{1}$ comes from the spin connection.

In what follows we assume the magnetic field configuration corresponding to
a magnetic flux located in the region $r<a$. This will be implemented by
considering the vector potential in the exterior region, $r>a$, as follows%
\begin{equation}
A_{\mu }=(0,0,A),  \label{Amu}
\end{equation}%
In Eq. (\ref{Amu}), $A_{2}=A$ is the covariant component of the vector
potential in the coordinates $(t,r,\phi )$. For the so called physical
azimuthal component one has $A_{\phi }=-A/r$. The quantity $A$ is related to
the magnetic flux $\Phi $ by the formula $A=-\Phi /\phi _{0}$. Though the
magnetic field strength, corresponding to (\ref{Amu}), vanishes outside the
flux, the non-trivial topology of the background spacetime leads to
Aharonov-Bohm-like effects on physical observables. In particular, as it
will be seen below, the VEV of the fermionic current depends on the
fractional part of the ratio of $\Phi $ by the quantum flux, $2\pi /e$.

Decomposing the spinor into upper and lower components, $\varphi _{+}$ and $%
\varphi _{-}$, respectively, from Eq. (\ref{Direq3}) we get the following
equations
\begin{equation}
\left( \partial _{0}\pm im\right) \varphi _{\pm }\pm ie^{\mp iq\phi }\Big[%
\partial _{1}+\frac{1-q}{2r}\mp \frac{i}{r}(\partial _{2}+ieA)\Big]\varphi
_{\mp }=0.  \label{DirEqphi}
\end{equation}%
From here we find the second-order differential equation for the separate
components:
\begin{equation}
\left( \partial _{0}^{2}-\partial _{1}^{2}-\frac{1}{r}\partial _{1}-\frac{1}{%
r^{2}}\partial _{2}^{2}\mp 2i\frac{c_{\pm }}{r^{2}}\partial _{2}+\frac{%
c_{\pm }^{2}}{r^{2}}+m^{2}\right) \varphi _{\pm }=0,  \label{DireqPhi1}
\end{equation}%
with the notations $c_{\pm }=(q-1)/2\pm eA$.

For the positive energy solutions, the dependence on the time and angle
coordinates is in the form $e^{-iEt+iqn\phi }$ with $E>0$ and $n=0,\pm 1,\pm
2,\ldots $. Now, from Eq. (\ref{DireqPhi1}), for the radial function we
obtain the Bessel equation with the solution in the region $r>a$:
\begin{equation}
\varphi _{+}=Z_{|\lambda _{n}|}(\gamma r)e^{iqn\phi -iEt},  \label{phiext}
\end{equation}%
where $\gamma \geqslant 0$,%
\begin{equation}
E=\sqrt{\gamma ^{2}+m^{2}},\;\lambda _{n}=q(n+\alpha +1/2)-1/2,
\label{gamma}
\end{equation}%
and%
\begin{equation}
\alpha =eA/q=-e\Phi /2\pi .  \label{alfatilde}
\end{equation}%
In Eq. (\ref{phiext}),%
\begin{equation}
Z_{|\lambda _{n}|}(\gamma r)=c_{1}J_{|\lambda _{n}|}(\gamma
r)+c_{2}Y_{|\lambda _{n}|}(\gamma r),  \label{Zsig}
\end{equation}%
with $J_{\nu }(x)$ and $Y_{\nu }(x)$ being the Bessel and Neumann functions.
Note that in Eq. (\ref{alfatilde}) the parameter $\alpha $ is the magnetic
flux measured in units of the flux quantum $\Phi _{0}=2\pi /e$.

The lower component of the spinor, $\varphi _{-}$, is found from Eq. (\ref%
{DirEqphi}) and for the positive energy eigenspinors we get%
\begin{equation}
\psi _{\gamma n}^{(+)}(x)=e^{iqn\phi -iEt}\left(
\begin{array}{c}
Z_{|\lambda _{n}|}(\gamma r) \\
\epsilon _{\lambda _{n}}\frac{\gamma e^{iq\phi }}{E+m}Z_{|\lambda
_{n}|+\epsilon _{\lambda _{n}}}(\gamma r)%
\end{array}%
\right) ,  \label{psisigpl}
\end{equation}%
where $\epsilon _{\lambda _{n}}=1$ for $\lambda _{n}\geqslant 0$ and $%
\epsilon _{\lambda _{n}}=-1$ for $\lambda _{n}<0$. From the boundary
condition (\ref{BCMIT}) with $n_{\mu }=-\delta _{\mu }^{1}$ we find%
\begin{equation}
Z_{|\lambda _{n}|}(\gamma a)+\frac{\epsilon _{\lambda _{n}}\gamma }{E+m}%
Z_{|\lambda _{n}|+\epsilon _{\lambda _{n}}}(\gamma a)=0.  \label{BCext}
\end{equation}%
This condition relates the coefficients $c_{1}$ and $c_{2}$ in the linear
combination (\ref{Zsig}):%
\begin{equation}
\frac{c_{2}}{c_{1}}=-\frac{\bar{J}_{|\lambda _{n}|}^{(-)}(\gamma a)}{\bar{Y}%
_{|\lambda _{n}|}^{(-)}(\gamma a)}.  \label{c21}
\end{equation}%
Here and in what follows we use the notations (the notation with the upper
sign is employed below)%
\begin{equation}
\bar{f}^{(\pm )}(z)=zf^{\prime }(z)+(\pm \sqrt{z^{2}+\mu ^{2}}\pm \mu
-\lambda _{n})f(z),\;\mu =ma,  \label{barnot2}
\end{equation}%
for a given function $f(z)$.

Hence, outside a circular boundary the positive energy eigenspinors are
presented in the form%
\begin{equation}
\psi _{\gamma n}^{(+)}(x)=c_{0}e^{iqn\phi -iEt}\left(
\begin{array}{c}
g_{|\lambda _{n}|,|\lambda _{n}|}(\gamma a,\gamma r) \\
\epsilon _{\lambda _{n}}\frac{\gamma e^{iq\phi }}{E+m}g_{|\lambda
_{n}|,|\lambda _{n}|+\epsilon _{\lambda _{n}}}(\gamma a,\gamma r)%
\end{array}%
\right) ,  \label{psisig+}
\end{equation}%
where%
\begin{equation}
g_{\nu ,\rho }(x,y)=\bar{Y}_{\nu }^{(-)}(x)J_{\rho }(y)-\bar{J}_{\nu
}^{(-)}(x)Y_{\rho }(y).  \label{gsig}
\end{equation}%
Using the properties of the Bessel functions it can be seen that
\begin{equation}
g_{\nu ,\nu }(x,y)=g_{-\nu ,-\nu }(x,y),\;g_{\nu ,\nu +1}(x,y)=-g_{-\nu
,-\nu -1}(x,y).  \label{gsig1}
\end{equation}%
Note that the spinor (\ref{psisig+}) is an eigenfunction of the operator $%
\widehat{J}=-(i/q)\partial _{\phi }+\sigma _{3}/2$, with the eigenvalue $%
j=n+1/2$, i.e.,%
\begin{equation}
\widehat{J}\psi _{\gamma n}^{(+)}(x)=j\psi _{\gamma n}^{(+)}(x),\;j=n+1/2.
\label{Jmom}
\end{equation}

The coefficient $c_{0}$ in Eq. (\ref{psisig+}) is determined from the
orthonormalization condition for the eigenspinors:
\begin{equation}
\int_{a}^{\infty }dr\int_{0}^{\phi _{0}}d\phi \,r\psi _{\gamma
n}^{(+)\dagger }(x)\psi _{\gamma ^{\prime }n^{\prime }}^{(+)}(x)=\delta
(\gamma -\gamma ^{\prime })\delta _{nn^{\prime }}\ .  \label{ortcon}
\end{equation}%
The integral over $r$ is divergent when $\gamma ^{\prime }=\gamma $ and,
hence, the main contribution comes from the upper limit of the integration.
In this case, we can replace the Bessel and Neumann functions, having in the
arguments the radial coordinate $r$, by the corresponding asymptotic
expressions for large values of their argument. In this way, for the
normalization coefficient we find,%
\begin{equation}
c_{0}^{2}=\frac{2E\gamma }{\phi _{0}(E+m)}\left[ \bar{J}_{|\lambda
_{n}|}^{(-)2}(\gamma a)+\bar{Y}_{|\lambda _{n}|}^{(-)2}(\gamma a)\right]
^{-1}.  \label{c0}
\end{equation}%
The negative energy eigenspinors are constructed in a similar way and they
are given by the expression
\begin{equation}
\psi _{\gamma n}^{(-)}(x)=c_{0}e^{-iqn\phi +iEt}\left(
\begin{array}{c}
\epsilon _{\lambda _{n}}\frac{\gamma e^{-iq\phi }}{E+m}g_{|\lambda
_{n}|,|\lambda _{n}|+\epsilon _{\lambda _{n}}}(\gamma a,\gamma r) \\
g_{|\lambda _{n}|,|\lambda _{n}|}(\gamma a,\gamma r)%
\end{array}%
\right) ,  \label{psisig-}
\end{equation}%
with the same normalization coefficient defined by Eq. (\ref{c0}). Note that
the positive and negative energy eigenspinors are related by the charge
conjugation which can be written as $\psi _{\gamma n}^{(-)}=\sigma _{1}\psi
_{\gamma n}^{(+)\ast }$, where the asterisk means complex conjugate.

We can generalize the eigenspinors given above for a more general situation
where the spinor field $\psi $ obeys quasiperiodic boundary condition along
the azimuthal direction%
\begin{equation}
\psi (t,r,\phi +\phi _{0})=e^{2\pi i\chi }\psi (t,r,\phi ),  \label{PerBC}
\end{equation}%
with a constant parameter $\chi $, $|\chi |\leqslant 1/2$. With this
condition, the exponential factor in the expressions for the eigenspinors
has the form $e^{\pm iq(n+\chi )\phi \mp iEt}$ for the positive and negative
energy modes (upper and lower signs respectively). The corresponding
expressions for the eigenfunctions are obtained from those given above with
the parameter $\alpha $ defined by
\begin{equation}
\alpha =\chi -e\Phi /2\pi .  \label{Replace}
\end{equation}%
The same replacement generalizes the expressions for the VEVs of the
fermionic current, given below, for the case of a field with periodicity
condition (\ref{PerBC}). The property, that the VEVs depend on the phase $%
\chi $ and on the magnetic flux in the combination (\ref{Replace}), can also
be seen by the gauge transformation $A_{\mu }=A_{\mu }^{\prime }+\partial
_{\mu }\Lambda (x)$, $\psi (x)=\psi ^{\prime }(x)e^{-ie\Lambda (x)}$, with
the function $\Lambda (x)=A_{\mu }x^{\mu }$. The new function $\psi ^{\prime
}(x)$ satisfies the Dirac equation with $A_{\mu }^{\prime }=0$ and the
quasiperiodicity condition similar to (\ref{PerBC}) with the replacement $%
\chi \rightarrow \chi ^{\prime }=\chi -e\Phi /2\pi $.

\section{Fermionic current in a boundary-free conical space}

\label{sec:BoundFree}

Before considering the fermionic current in the region outside a circular
boundary, in this section we study the case of a boundary-free conical space
with an infinitesimally thin magnetic flux placed at the apex of the cone.
The corresponding vector potential is given by Eq. (\ref{Amu}) for $r>0$. As
it is well known, the theory of von Neumann deficiency indices leads to a
one-parameter family of allowed boundary conditions in the background of an
Aharonov-Bohm gauge field \cite{Sous89}. In this paper, we consider a
special case of boundary conditions at the cone apex, when the MIT bag
boundary condition is imposed at a finite radius, which is then taken to
zero (note that similar approach, with the Atiyah-Patodi-Singer type
nonlocal boundary conditions, has been used in Refs. \cite{Bene00} for a
magnetic flux in Minkowski spacetime). The VEVs of the fermionic current for
other boundary conditions on the cone apex are evaluated in a way similar to
that described below. The contribution of the regular modes is the same for
all boundary conditions and the results differ by the parts related to the
irregular modes.

\subsection{Eigenspinors}

In order to clarify the structure of the eigenspinors in a boundary-free
conical space, we consider the limit $a\rightarrow 0$ for Eqs. (\ref{psisig+}%
) and (\ref{psisig-}). In this limit, using the asymptotic formulae for the
Bessel functions for small values of the arguments, for the modes with $%
j\neq -\alpha $ we find%
\begin{eqnarray}
\psi _{(0)\gamma j}^{(+)}(x) &=&c_{0}^{(0)}e^{iqj\phi -iEt}\left(
\begin{array}{c}
J_{\beta _{j}}(\gamma r)e^{-iq\phi /2} \\
\frac{\gamma \epsilon _{j}e^{iq\phi /2}}{E+m}J_{\beta _{j}+\epsilon
_{j}}(\gamma r)%
\end{array}%
\right) ,  \notag \\
\psi _{(0)\gamma j}^{(-)}(x) &=&c_{0}^{(0)}e^{-iqj\phi +iEt}\left(
\begin{array}{c}
\frac{\gamma \epsilon _{j}e^{-iq\phi /2}}{E+m}J_{\beta _{j}+\epsilon
_{j}}(\gamma r) \\
J_{\beta _{j}}(\gamma r)e^{iq\phi /2}%
\end{array}%
\right) ,  \label{psi0}
\end{eqnarray}%
where%
\begin{equation}
\beta _{j}=q|j+\alpha |-\epsilon _{j}/2.  \label{jbetj}
\end{equation}%
Note that one has $\epsilon _{j}\beta _{j}=\lambda _{n}$. In Eqs. (\ref{psi0}%
) and (\ref{jbetj}), we have defined%
\begin{equation}
\epsilon _{j}=\left\{
\begin{array}{cc}
1, & \;j>-\alpha \\
-1, & \;j<-\alpha%
\end{array}%
\right. ,  \label{epsj}
\end{equation}%
and the normalization coefficient is given by the expression%
\begin{equation}
c_{0}^{(0)2}=\frac{\gamma }{\phi _{0}}\frac{E+m}{2E}.  \label{c00}
\end{equation}

In the case when $\alpha =N+1/2$, with $N$ being an integer number, the
eigenspinors with $j\neq -\alpha $ are still given by Eqs. (\ref{psi0}). The
eigenspinors for the mode with $j=-\alpha $, obtained from Eqs. (\ref%
{psisig+}) and (\ref{psisig-}) in the limit $a\rightarrow 0$, have the form (%
\ref{psi0}) with the replacements%
\begin{eqnarray}
J_{\beta _{j}}(z) &\rightarrow &(E+m)J_{1/2}(z)-\gamma Y_{1/2}(z),  \notag \\
J_{\beta _{j}+\epsilon _{j}}(z) &\rightarrow &(E+m)J_{-1/2}(z)-\gamma
Y_{-1/2}(z),  \label{jeqalfa}
\end{eqnarray}%
and $\epsilon _{j}=-1$. The corresponding normalization coefficient is
defined as $c_{0}^{(0)}=(2E)^{-1}\sqrt{\gamma /\phi _{0}}$. Taking into
account the expressions for the cylinder functions with the orders $\pm 1/2$%
, the negative energy eigenspinors in this case are written as%
\begin{equation}
\psi _{(0)\gamma ,-\alpha }^{(-)}(x)=\left( \frac{E+m}{\pi \phi _{0}rE}%
\right) ^{1/2}e^{iq\alpha \phi +iEt}\left(
\begin{array}{c}
\frac{\gamma e^{-iq\phi /2}}{E+m}\sin (\gamma r-\gamma _{0}) \\
e^{iq\phi /2}\cos (\gamma r-\gamma _{0})%
\end{array}%
\right) ,  \label{psibetSp}
\end{equation}%
where $\gamma _{0}=\arccos [\sqrt{(E-m)/2E}]$. Note that the eigenspinors
obtained from (\ref{psi0}) in the limits $\alpha \rightarrow (N+1/2)^{\pm }$
do not coincide with (\ref{psibetSp}). For the limit from below (above), $%
\alpha \rightarrow (N+1/2)^{-}$ ($\alpha \rightarrow (N+1/2)^{+}$), the
eigenspinors are given by Eq. (\ref{psibetSp}) with the replacement $\gamma
_{0}\rightarrow \pi /2$ ($\gamma _{0}\rightarrow 0$). Hence, for the bag
boundary condition on the cone apex, the eigenspinors in the boundary-free
geometry are discontinuous at points $\alpha =N+1/2$. Notice that, in the
presence of the circular boundary, the eigenspinors in the region outside
the boundary, given by Eqs. (\ref{psisig+}) and (\ref{psisig-}), are
continuous.

In general, the fermionic modes in background of the magnetic vortex are
divided into two classes, regular and irregular (square integrable) ones. In
the problem under consideration, for given $q$ and $\alpha $, the irregular
mode corresponds to the value of $j$ for which $q|j+\alpha |<1/2$. If we
present the parameter $\alpha $ in the form%
\begin{equation}
\alpha =\alpha _{0}+n_{0},\;|\alpha _{0}|<1/2,  \label{alf0}
\end{equation}%
being $n_{0}$ an integer number, then the irregular mode is present if $%
|\alpha _{0}|>(1-1/q)/2$. This mode corresponds to $j=-n_{0}-$sgn$(\alpha
_{0})/2$. Note that, in a conical space, under the condition
\begin{equation}
|\alpha _{0}|\leqslant (1-1/q)/2,  \label{condalf0}
\end{equation}%
there are no square integrable irregular modes. As we have already
mentioned, there is a one-parameter family of allowed boundary conditions
for irregular modes, parametrized with the angle $\theta $, $0\leqslant
\theta <2\pi $ (see Ref. \cite{Sous89}). For $|\alpha _{0}|<1/2$, the
boundary condition, used in deriving eigenspinors (\ref{psi0}), corresponds
to $\theta =3\pi /2$. If $\alpha $ is a half-integer, the irregular mode
corresponds to $j=-\alpha $ and for the corresponding boundary condition one
has $\theta =0$. Note that in both cases there are no bound states.

\subsection{Vacuum expectation value of the fermionic current}

The VEV of the fermionic current, $j^{\mu }(x)=e\bar{\psi}\gamma ^{\mu }\psi
$, can be evaluated by using the mode sum formula%
\begin{equation}
\langle j^{\nu }(x)\rangle =e\sum_{j}\int_{0}^{\infty }d\gamma \,\bar{\psi}%
_{\gamma j}^{(-)}(x)\gamma ^{\nu }\psi _{\gamma j}^{(-)}(x),  \label{FCMode}
\end{equation}
where $\sum_{j}$ means the summation over $j=\pm 1/2,\pm 3/2,\ldots $. In
this section we consider this VEV for a conical space in the absence of
boundaries. The corresponding quantities will be denoted by subscript 0. For
the geometry under consideration, the eigenspinors are given by expressions (%
\ref{psi0}). Substituting them into Eq. (\ref{FCMode}) one finds%
\begin{eqnarray}
\langle j^{0}(x)\rangle _{0} &=&\frac{eq}{4\pi }\sum_{j}\int_{0}^{\infty
}d\gamma \frac{\gamma }{E}\left[ (E-m)J_{\beta _{j}+\epsilon
_{j}}^{2}(\gamma r)+(E+m)J_{\beta _{j}}^{2}(\gamma r)\right] ,  \notag \\
\langle j^{2}(x)\rangle _{0} &=&\frac{eq}{2\pi r}\sum_{j}\epsilon
_{j}\int_{0}^{\infty }d\gamma \frac{\gamma ^{2}}{E}J_{\beta _{j}}(\gamma
r)J_{\beta _{j}+\epsilon _{j}}(\gamma r),  \label{j020}
\end{eqnarray}%
and the VEV of the radial component vanishes, $\langle j^{1}(x)\rangle
_{0}=0 $. In deriving Eqs. (\ref{j020}), we have assumed that the parameter $%
\alpha $ is not a half-integer. When $\alpha $ is equal to a half-integer,
the contribution of the mode with $j=-\alpha $ should be evaluated by using
eigenspinors (\ref{psibetSp}). The contribution for all other $j$ is still
given by Eqs. (\ref{j020}). As it will be shown below, both these
contributions are separately zero and for half-integer values of $\alpha $
the renormalized VEV of the fermionic current vanishes.

In order to regularize expressions (\ref{j020}) we introduce a cutoff
function $e^{-s\gamma ^{2}}$ with the cutoff parameter $s>0$. At the end of
calculations the limit $s\rightarrow 0$ is taken. First let us consider the
charge density. The corresponding regularized expectation value is presented
in the form%
\begin{eqnarray}
\langle j^{0}(x)\rangle _{0,\text{reg}} &=&\frac{eqm}{4\pi }%
\sum_{j}\int_{0}^{\infty }d\gamma \,\frac{\gamma e^{-s\gamma ^{2}}}{\sqrt{%
\gamma ^{2}+m^{2}}}\left[ J_{\beta _{j}}^{2}(\gamma r)-J_{\beta
_{j}+\epsilon _{j}}^{2}(\gamma r)\right]  \notag \\
&&+\frac{eq}{4\pi }\sum_{j}\int_{0}^{\infty }d\gamma \,\gamma e^{-s\gamma
^{2}}\left[ J_{\beta _{j}}^{2}(\gamma r)+J_{\beta _{j}+\epsilon
_{j}}^{2}(\gamma r)\right] .  \label{j00reg}
\end{eqnarray}%
Using the representation%
\begin{equation}
\frac{1}{\sqrt{\gamma ^{2}+m^{2}}}=\frac{2}{\sqrt{\pi }}\int_{0}^{\infty
}dte^{-(\gamma ^{2}+m^{2})t^{2}},  \label{repres}
\end{equation}%
we change the order of integrations in the first term of the right-hand side
in Eq. (\ref{j00reg}) and use the formula \cite{Prud86}
\begin{equation}
\int_{0}^{\infty }d\gamma \,\gamma e^{-s\gamma ^{2}}J_{\beta }^{2}(\gamma r)=%
\frac{1}{2s}e^{-r^{2}/2s}I_{\beta }(r^{2}/2s),  \label{intform}
\end{equation}%
with $I_{\beta }(z)$ being the modified Bessel function. The second term on
the right of Eq. (\ref{j00reg}) is directly evaluated using Eq. (\ref%
{intform}). As a result, we get the following integral representation for
the regularized charge density:%
\begin{eqnarray}
\langle j^{0}(x)\rangle _{0,\text{reg}} &=&\frac{eqme^{m^{2}s}}{2(2\pi
)^{3/2}}\sum_{j}\int_{0}^{r^{2}/2s}dz\frac{z^{-1/2}e^{-m^{2}r^{2}/2z}}{\sqrt{%
r^{2}-2zs}}e^{-z}\left[ I_{\beta _{j}}(z)-I_{\beta _{j}+\epsilon _{j}}(z)%
\right]  \notag \\
&&+\frac{eqe^{-r^{2}/2s}}{8\pi s}\sum_{j}\left[ I_{\beta
_{j}}(r^{2}/2s)+I_{\beta _{j}+\epsilon _{j}}(r^{2}/2s)\right] .
\label{j00reg1}
\end{eqnarray}%
The renormalization procedure for this expression is described below.

Now we turn to the azimuthal component of the fermionic current. Using the
relation%
\begin{equation}
zJ_{\beta _{j}+\epsilon _{j}}(z)=\beta _{j}J_{\beta _{j}}(z)-\epsilon
_{j}zJ_{\beta _{j}}^{\prime }(z),  \label{relBess}
\end{equation}%
we write the corresponding regularized expression in the form%
\begin{equation}
\langle j^{2}(x)\rangle _{0,\text{reg}}=\frac{eq}{2\pi r^{2}}\sum_{j}\left(
\epsilon _{j}\beta _{j}-r\partial _{r}/2\right) \int_{0}^{\infty }d\gamma
\,\gamma \frac{e^{-s\gamma ^{2}}J_{\beta _{j}}^{2}(xr)}{\sqrt{\gamma
^{2}+m^{2}}}.  \label{j02reg}
\end{equation}%
In a way similar to that we have used for the first term in the right-hand
side of Eq. (\ref{j00reg}), the azimuthal current is presented in the form%
\begin{equation}
\langle j^{2}(x)\rangle _{0,\text{reg}}=\frac{eqe^{m^{2}s}}{(2\pi
)^{3/2}r^{2}}\sum_{j}\int_{0}^{r^{2}/2s}dz\frac{z^{1/2}e^{-m^{2}r^{2}/2z}}{%
\sqrt{r^{2}-2zs}}e^{-z}\left[ I_{\beta _{j}}(z)-I_{\beta _{j}+\epsilon
_{j}}(z)\right] .  \label{j02reg1}
\end{equation}%
In deriving this representation we employed the relation%
\begin{equation}
\left( \epsilon _{j}\beta _{j}-r\partial _{r}/2\right) e^{-r^{2}y}I_{\beta
_{j}}(r^{2}y)=ze^{-z}\left[ I_{\beta _{j}}(z)-I_{\beta _{j}+\epsilon _{j}}(z)%
\right] _{z=r^{2}y},  \label{relBesMod}
\end{equation}%
for the modified Bessel function.

The expressions of the regularized VEVs for both charge density and
azimuthal current are expressed in terms of the series%
\begin{equation}
\mathcal{I}(q,\alpha ,z)=\sum_{j}I_{\beta _{j}}(z).  \label{seriesI0}
\end{equation}%
If we present the parameter $\alpha $ related to the magnetic flux as (\ref%
{alf0}), then Eq. (\ref{seriesI0}) is written in the form%
\begin{equation}
\mathcal{I}(q,\alpha ,z)=\sum_{n=0}^{\infty }\left[ I_{q(n+\alpha
_{0}+1/2)-1/2}(z)+I_{q(n-\alpha _{0}+1/2)+1/2}(z)\right] ,  \label{seriesI1}
\end{equation}%
which explicitly shows the independence of the series on $n_{0}$. Note that
for the second series appearing in the expressions for the VEVs of the
fermionic current we have
\begin{equation}
\sum_{j}I_{\beta _{j}+\epsilon _{j}}(z)=\mathcal{I}(q,-\alpha _{0},z).
\label{seriesI2}
\end{equation}%
We conclude that the VEVs of the fermionic current depend on $\alpha _{0}$
alone and, hence, these VEVs are periodic functions of $\alpha $ with period
1.

When the parameter $\alpha $ is equal to a half-integer, that means $|\alpha
_{0}|=1/2$, the contribution to the VEVs from the modes with $j\neq -\alpha $
is still given by Eqs. (\ref{j00reg1}) and (\ref{j02reg1}). It is easily
seen that for the case under consideration $\sum_{j\neq -\alpha }\left[
I_{\beta _{j}}(x)-I_{\beta _{j}+\epsilon _{j}}(x)\right] =0$. The
contribution of the mode $j=-\alpha $ is evaluated by using eigenspinors (%
\ref{psibetSp}). A simple evaluation shows that this contribution vanishes
as well. As regards the second term on the right-hand side of Eq. (\ref%
{j00reg1}), below it will be shown that this term does not contribute to the
renormalized VEV of the charge density. Hence, the renormalized VEVs for
both charge density and azimuthal current vanish in the case when the
parameter $\alpha $ is equal to a half-integer. Note that in the limit $%
\alpha _{0}\rightarrow \pm 1/2$, $|\alpha _{0}|<1/2$, one has%
\begin{equation}
\lim_{\alpha _{0}\rightarrow \pm 1/2}\sum_{\delta =\pm 1}\delta \mathcal{I}%
(q,\delta \alpha _{0},z)=\mp \sqrt{2/\pi z}e^{-z},  \label{RelLim}
\end{equation}%
and the expressions for the regularized VEVs are discontinuous at $\alpha
_{0}=\pm 1/2$.

\subsection{Renormalized VEV in a special case}

Before further considering the fermionic current for the general case of the
parameters characterizing the conical structure and the magnetic flux, we
study a special case, which allows us to obtain simple expressions. It has
been shown in \cite{Davi88,Smit89,Sour92} that when the parameter $q$ is an
integer number, the scalar Green function in four-dimensional cosmic string
spacetime can be expressed as a sum of $q$ images of the Minkowski spacetime
function. Also, recently the image method was used in \cite{Beze06} to
provide closed expressions for the massive scalar Green functions in a
higher-dimensional cosmic string spacetime. The mathematical reason for the
use of the image method in these applications is because the order of the
modified Bessel functions that appear in the expressions for the VEVs
becomes an integer number. As we have seen, for the fermionic case the order
of the Bessel function depends, besides on the integer angular quantum
number $n=j-1/2$, also on the factor $(q-1)/(2q)$ which comes from the spin
connection. However, considering a charged fermionic field in the presence
of a magnetic flux, an additional term will be present, the factor $\alpha $%
. In the special case with $q$ being an integer and
\begin{equation}
\alpha =1/2q-1/2,  \label{alphaSpecial}
\end{equation}%
the orders of the modified Bessel functions in Eqs. (\ref{j00reg1}) and (\ref%
{j02reg1}) become integer numbers: $\beta _{j}=q|n|$, $j=n+1/2$. In this
case the series over $n$ is summarized explicitly by using the formula \cite%
{Prud86}%
\begin{equation}
\sideset{}{'}{\sum}_{n=0}^{\infty }I_{qn}(x)=\frac{1}{2q}%
\sum_{k=0}^{q-1}e^{x\cos (2\pi k/q)},  \label{SerSp}
\end{equation}%
where the prime on the summation sign means that the term $n=0$ should be
halved.

By making use of Eq. (\ref{SerSp}), for the regularized VEV of charge
density we find%
\begin{eqnarray}
\langle j^{0}(x)\rangle _{0,\text{reg}} &=&\frac{eme^{m^{2}s}}{(2\pi )^{3/2}}%
\sum_{k=1}^{q-1}\sin ^{2}(\pi k/q)\int_{0}^{r^{2}/2s}dz\frac{%
z^{-1/2}e^{-m^{2}r^{2}/2z}}{\sqrt{r^{2}-2zs}}e^{-2z\sin ^{2}(\pi k/q)}
\notag \\
&&+\frac{e}{4\pi s}\sum_{k=0}^{q-1}\cos ^{2}(\pi k/q)e^{-2(r^{2}/2s)\sin
^{2}(\pi k/q)}.  \label{j00Spreg}
\end{eqnarray}%
In the limit $s\rightarrow 0$, the only divergent contribution to the
right-hand side comes from the term $k=0$. This term does not depend on the
parameter $q$ and is the same as in the Minkowski spacetime in the absence
of the magnetic flux. Subtracting the $k=0$ term, then we take the limit $%
s\rightarrow 0$. The integral is expressed in terms of the Macdonald
function $K_{1/2}(2mr\sin (\pi k/q))$ and for the renormalized charge
density we find:%
\begin{equation}
\langle j^{0}(x)\rangle _{0,\text{ren}}=\frac{em}{4\pi r}\sum_{k=1}^{q-1}%
\sin (\pi k/q)e^{-2mr\sin (\pi k/q)}.  \label{j00Spren}
\end{equation}%
Note that the second term on the right-hand side of Eq. (\ref{j00reg1}) does
not contribute to the renormalized VEV. Expression (\ref{j00Spren})
coincides with the result of Ref. \cite{Beze10} obtained by using the Green
function approach.

\begin{figure}[tbph]
\begin{center}
\begin{tabular}{cc}
\epsfig{figure=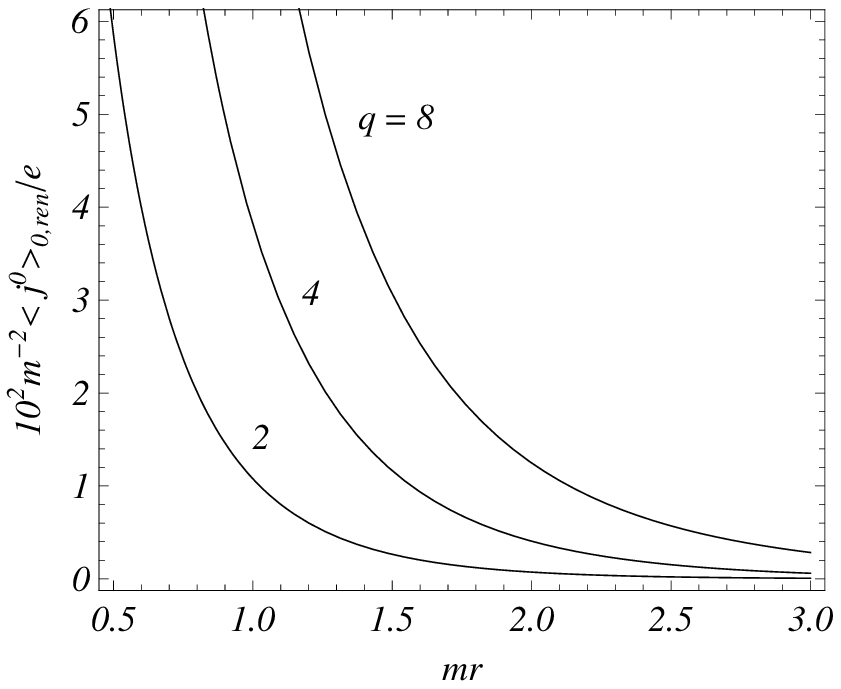,width=7.cm,height=6.cm} & \quad %
\epsfig{figure=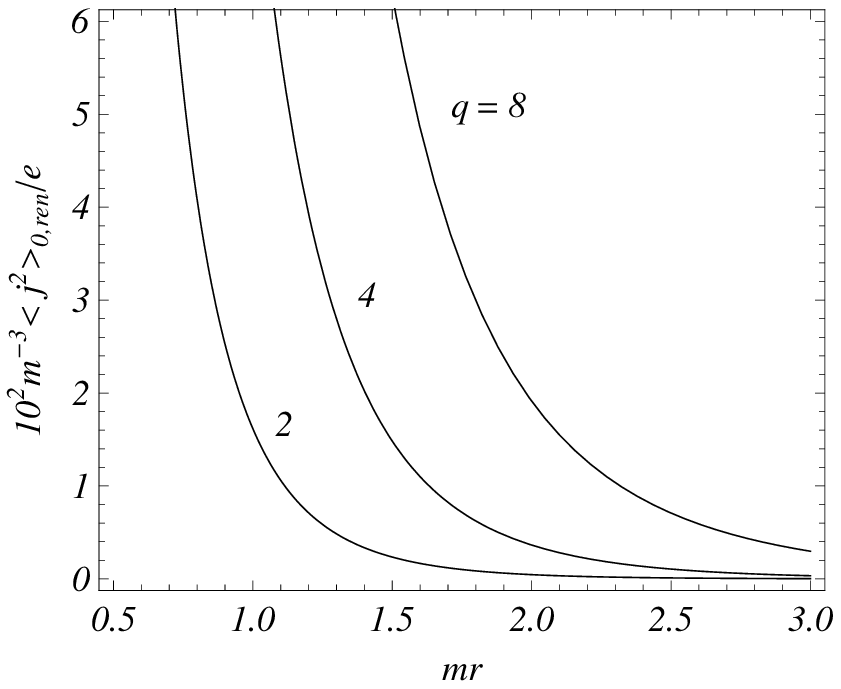,width=7.cm,height=6.cm}%
\end{tabular}%
\end{center}
\caption{VEVs of the charge density (left plot) and the azimuthal current
(right plot) in a boundary-free conical space, as functions of the parameter
$mr$ for the special case of integer values of $q$ with the magnetic flux
defined by Eq. (\protect\ref{alphaSpecial}).}
\label{fig1}
\end{figure}

In a similar way, using (\ref{SerSp}), for the regularized VEV\ of the
azimuthal current we get the formula%
\begin{equation}
\langle j^{2}(x)\rangle _{0,\text{reg}}=\frac{e}{\pi r^{2}}\frac{e^{m^{2}s}}{%
\sqrt{2\pi }}\sum_{k=1}^{q-1}\sin ^{2}(\pi k/q)\int_{0}^{r^{2}/2s}dz\frac{%
z^{1/2}e^{-m^{2}r^{2}/2z}}{\sqrt{r^{2}-2zs}}e^{-2z\sin ^{2}(\pi k/q)}.
\label{j02Spreg}
\end{equation}%
This expression is finite in the limit $s\rightarrow 0$ and for the
renormalized VEV one finds%
\begin{equation}
\langle j^{2}(x)\rangle _{0,\text{ren}}=\frac{e}{8\pi r^{3}}\sum_{k=1}^{q-1}%
\frac{1+2mr\sin (\pi k/q)}{\sin (\pi k/q)}e^{-2mr\sin (\pi k/q)}.
\label{j02Spren}
\end{equation}%
After the coordinate transformation $\phi ^{\prime }=q\phi $, this
expression coincides with the result given in Ref. \cite{Beze10}. Note that
for both charge density and azimuthal current one has $\langle j^{\nu
}(x)\rangle _{0,\text{ren}}/e\geqslant 0$. As expected the renormalized
current densities decay exponentially at distances larger than the Compton
wavelength of the fermionic particle. In Fig. \ref{fig1} the VEVs of charge
density and azimuthal current are plotted versus $mr$ for different values
of $q$. The corresponding values of the parameter $\alpha $ are found from
Eq. (\ref{alphaSpecial}).

\subsection{General case}

Now we turn to the general case for the parameters $q$ and $\alpha $. As it
follows from Eqs. (\ref{j00reg1}) and (\ref{j02reg1}), we need to evaluate
the integrals%
\begin{equation}
\int_{0}^{r^{2}/2s}dx\frac{x^{\pm 1/2}e^{-m^{2}r^{2}/2x}}{\sqrt{r^{2}-2xs}}%
e^{-x}\left[ \mathcal{I}(q,\alpha _{0},x)-\mathcal{I}(q,-\alpha _{0},x)%
\right] ,  \label{Integrals}
\end{equation}%
in the limit $s\rightarrow 0$, with the function $\mathcal{I}(q,\alpha
_{0},x)$ defined by Eq. (\ref{seriesI0}). Two alternative integral
representations for this series are given in Appendix \ref{sec:IntRep}. In
order to provide an integral representation for the VEV\ of fermionic
current, we first consider the representation (\ref{seriesI3}). In the limit
$s\rightarrow 0$, the only divergent contributions to the integrals in Eq. (%
\ref{Integrals}) with separate functions $\mathcal{I}(q,\pm \alpha _{0},x)$
come from the first term in the right-hand side of Eq. (\ref{seriesI3}).
This term does not depend on $\alpha _{0}$ and, consequently, it is
cancelled in the evaluation of the integral in Eq. (\ref{Integrals}).

Hence, we see that the regularized expression (\ref{j02reg1}) for the VEV\
of azimuthal current is finite in the limit $s\rightarrow 0$. For the
corresponding renormalized quantity we find%
\begin{equation}
\langle j^{2}(x)\rangle _{0,\text{ren}}=\frac{eqr^{-3}}{(2\pi )^{3/2}}%
\int_{0}^{\infty }dz\,z^{1/2}e^{-m^{2}r^{2}/2z-z}\sum_{\delta =\pm 1}\delta
\mathcal{I}(q,\delta \alpha _{0},z).  \label{j02ren1}
\end{equation}%
where the function $\mathcal{I}(q,\alpha _{0},z)$ is given by the integral
representation (\ref{seriesI3}). The VEV\ of the azimuthal current is a
periodical function of the magnetic flux with a period equal to magnetic
flux quantum $\Phi _{0}$. It is an odd function of the parameter $\alpha
_{0} $ defined by Eq. (\ref{alf0}). From Eq. (\ref{seriesI3}) it is seen
that the integrand in Eq. (\ref{j02ren1}) decays exponentially for large
values of $z$. Substituting the integral representation (\ref{seriesI3})
into Eq. (\ref{j02ren1}) and changing the order of integrations in the part
with the last term of Eq. (\ref{seriesI3}), the integral over $z$ is
expressed in terms of the Macdonald function $K_{3/2}(y)$.

As a result, for the renormalized VEV of the azimuthal component we find the
following expression%
\begin{eqnarray}
&& \langle j^{2}(x)\rangle _{0,\text{ren}} =\frac{e}{4\pi r^{3}}\Big\{%
\sum_{l=1}^{p}\frac{(-1)^{l}\sin (2\pi l\alpha _{0})}{\sin ^{2}(\pi l/q)}%
\frac{1+2mr\sin (\pi l/q)}{e^{2mr\sin (\pi l/q)}}  \notag \\
&& \qquad -\frac{q}{4\pi }\int_{0}^{\infty }dy\frac{\sum_{\delta =\pm
1}\delta f(q,\delta \alpha _{0},y)}{\cosh (qy)-\cos (q\pi )}\frac{1+2mr\cosh
(y/2)}{\cosh ^{3}(y/2)e^{2mr\cosh (y/2)}}\Big\},  \label{j02ren2}
\end{eqnarray}
where $p$ is an integer defined by $2p<q<2p+2$ and the function $f(q,\alpha
_{0},y)$ is given by Eq. (\ref{fqualf}). In the case $q=2p$, the additional
term
\begin{equation}
-(-1)^{q/2}\sin (\pi q\alpha _{0})e^{-2mr}(1/2+mr),  \label{AddTermj2}
\end{equation}%
should be added to the expression in the figure braces of Eq. (\ref{j02ren2}%
). For $1\leqslant q<2$, only the integral term remains. The difference of
the functions $f(q,\alpha _{0},y)$, appearing in the integrand, can also be
written in the form
\begin{equation}
\sum_{\delta =\pm 1}\delta f(q,\delta \alpha _{0},y)=2\cosh
(y/2)\sum_{\delta =\pm 1}\delta \cos \left[ q\pi \left( 1/2-\delta \alpha
_{0}\right) \right] \cosh \left[ q\left( 1/2+\delta \alpha _{0}\right) y%
\right] .  \label{Diff}
\end{equation}%
Note that for $q=2p$ the integrand in the last term of Eq. (\ref{j02ren2})
is finite at $y=0$.

\begin{figure}[tbph]
\begin{center}
\epsfig{figure=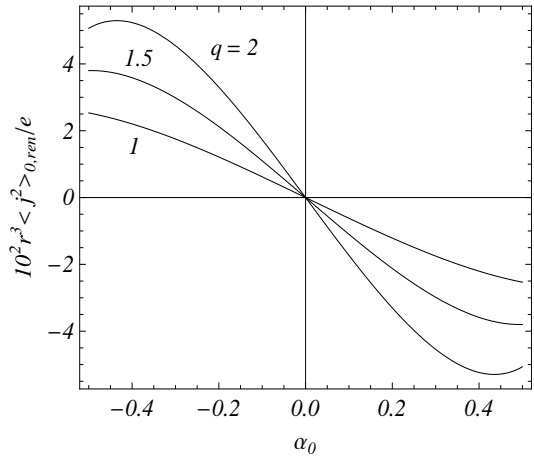,width=7.cm,height=6.cm}
\end{center}
\caption{The VEV of the azimuthal current in a boundary-free conical space
as a function of the magnetic flux for a massless fermionic field.}
\label{fig2}
\end{figure}

In the massless limit the expression in the figure braces of Eq. (\ref%
{j02ren2}) does not depend on the radial coordinate $r$ and the renormalized
VEV of the azimuthal current behaves as $1/r^{3}$. For a massive field, at
distances larger than the Compton wavelength of the spinor particle, $mr\gg
1 $, the VEV of the azimuthal current is suppressed by the factor $e^{-2mr}$
for $1\leqslant q<2$ and by the factor $e^{-2mr\sin (\pi /q)}$ for $%
q\geqslant 2$. In the limit $mr\ll 1$, the leading term in the corresponding
asymptotic expansion coincides with the VEV for a massless field. In Fig. %
\ref{fig2} we plot the VEV of the azimuthal current for a massless fermionic
field as a function of the magnetic flux for several values of the parameter
$q$ (numbers near the curves). In the limit $\alpha _{0}\rightarrow \pm 1/2$%
, $|\alpha _{0}|<1/2$, the VEV of the azimuthal current is obtained using
relation (\ref{RelLim}). From Eq. (\ref{j02ren1}) one finds:%
\begin{equation}
\lim_{\alpha _{0}\rightarrow \pm 1/2}\langle j^{2}(x)\rangle _{0,\text{ren}%
}=\mp \frac{eqm}{2\pi ^{2}r^{2}}K_{1}(2mr),  \label{Limj2}
\end{equation}%
where $K_{\nu }(z)$ is the Macdonald function.

Now we turn to the VEV of the charge density. In the corresponding
regularized expression (\ref{j00reg1}), the first term on the right-hand
side with the integral is finite in the limit $s\rightarrow 0$. The second
term is written in the form%
\begin{equation}
\frac{eqe^{-r^{2}/2s}}{8\pi s}\sum_{\delta =\pm 1}\mathcal{I}(q,\delta
\alpha _{0},r^{2}/2s).  \label{Term}
\end{equation}%
Taking into account Eq. (\ref{seriesI3}), we see that the only nonzero
contribution to this term comes from the first term in the right-hand side
of Eq. (\ref{seriesI3}). This contribution diverges and does not depend on
the parameters $q$ and $\alpha _{0}$. The divergence is the same as in the
Minkowski spacetime in the absence of the magnetic flux and is subtracted in
the renormalization procedure. As a result, for the renormalized VEV of the
charge density we get the formula%
\begin{equation}
\langle j^{0}(x)\rangle _{0,\text{ren}}=\frac{eqm}{2(2\pi )^{3/2}r}%
\int_{0}^{\infty }dz\,z^{-1/2}e^{-m^{2}r^{2}/2z-z}\sum_{\delta =\pm 1}\delta
\mathcal{I}(q,\delta \alpha _{0},z).  \label{j00ren1}
\end{equation}%
This VEV is a periodical function of the magnetic flux with a period equal
to the magnetic flux quantum. For $2p<q<2p+2$, with $p$ being an integer,
using the integral representation (\ref{seriesI3}), the renormalized VEV is
presented in the form%
\begin{eqnarray}
&& \langle j^{0}(x)\rangle _{0,\text{ren}} =\frac{em}{2\pi r}\Big\{%
\sum_{l=1}^{p}(-1)^{l}\sin (2\pi l\alpha _{0})e^{-2mr\sin (\pi l/q)}  \notag
\\
&&\qquad -\frac{q}{4\pi }\int_{0}^{\infty }dy\frac{\sum_{\delta =\pm
1}\delta f(q,\delta \alpha _{0},y)}{\cosh (qy)-\cos (q\pi )}\frac{%
e^{-2mr\cosh (y/2)}}{\cosh (y/2)}\Big\}.  \label{j00ren2}
\end{eqnarray}%
When $q=2p$, the additional term
\begin{equation}
-(-1)^{q/2}\sin (\pi q\alpha _{0})e^{-2mr}/2,  \label{AddTermj0}
\end{equation}%
should be added to the expression in the figure braces on the right-hand
side of Eq. (\ref{j02ren2}). As in the case of the azimuthal current, the
renormalized VEV (\ref{j00ren1}) is an odd function of the parameter $\alpha
_{0}$. Note that in a boundary-free conical space the charge density for a
massless field vanishes at points outside the magnetic flux. For a massive
field, at distances larger than the Compton wavelength, $mr\gg 1$, the
renormalized charge density is suppressed by the factor $e^{-2mr}$ for $%
1\leqslant q<2$ and by the factor $e^{-2mr\sin (\pi /q)}$ for $q\geqslant 2$.

Though the charge density given by Eq. (\ref{j00ren1}) diverges at $r=0$,
this divergence is integrable and the total fermionic charge%
\begin{equation}
Q=\phi _{0}\int_{0}^{\infty }dr\,r\langle j^{0}(x)\rangle _{0,\text{ren}}=%
\frac{e}{4}\int_{0}^{\infty }dx\,e^{-x}\sum_{\delta =\pm 1}\delta \mathcal{I}%
(q,\delta \alpha _{0},x),  \label{Q}
\end{equation}%
is finite. In the form given by Eq. (\ref{Q}), the integrals with $\delta =1$
and $\delta =-1$ diverge separately and we cannot change the order of the
integration an summation over $\delta $. In order to overcome this
difficulty, we write the integral as $\int_{0}^{\infty }dx\,e^{-x}\cdots
=\lim_{s\rightarrow 1^{+}}\int_{0}^{\infty }dx\,e^{-sx}\cdots $. With this
representation, evaluating the integrals with separate $\delta $ for $s>1$
and taking the limit $s\rightarrow 1$, one finds%
\begin{equation}
Q=-e\alpha _{0}/2.  \label{Q1}
\end{equation}%
This result for a conical space was previously obtained in Ref. \cite{Beze94}%
. As we see, the total charge does not depend on the angle deficit of the
conical space. This property is a consequence of the fact that the total
charge is a topologically invariant quantity depending only on the net flux.

In the absence of the angle deficit one has $q=1$ and the expressions for
the VEVs of the fermionic current are simplified to%
\begin{eqnarray}
\langle j^{0}(x)\rangle _{0,\text{ren}} &=&-\frac{em\sin \left( \pi \alpha
_{0}\right) }{2\pi ^{2}r}\int_{0}^{\infty }dz\,\frac{\cosh (2\alpha _{0}z)}{%
\cosh z}e^{-2mr\cosh z},  \notag \\
\langle j^{2}(x)\rangle _{0,\text{ren}} &=&-\frac{e\sin \left( \pi \alpha
_{0}\right) }{4\pi ^{2}r^{3}}\int_{0}^{\infty }dz\,\frac{\cosh (2\alpha
_{0}z)}{\cosh ^{3}z}\left( 1+2mr\cosh z\right) e^{-2mr\cosh z}.  \label{FCq1}
\end{eqnarray}%
Alternative expressions for the VEVs of the charge density and azimuthal
current in (2+1)-dimensional Minkowski spacetime in the presence of a
magnetic flux were given in Ref. \cite{Flek91} (see also \cite{Site99} for
the general case of a one-parameter family of boundary conditions at the
origin). We compare these expressions with the formulae obtained in the
present paper in Appendix \ref{sec:App2New}.

In Fig. \ref{fig3}, the VEVs of the charge density (left panel) and
azimuthal current (right panel) are plotted as functions of the magnetic
flux for a massive fermionic field in a conical space with $\phi _{0}=\pi $.
In the limit $\alpha _{0}\rightarrow \pm 1/2$, $|\alpha _{0}|<1/2$, for the
azimuthal current one has Eq. (\ref{Limj2}). For the charge density the
limiting values are given by
\begin{equation}
\lim_{\alpha _{0}\rightarrow \pm 1/2}\langle j^{0}(x)\rangle _{0,\text{ren}%
}=\mp \frac{eqm}{2\pi ^{2}r}K_{0}(2mr).  \label{Limj0}
\end{equation}%
Both charge density and azimuthal current exhibit the jump structure at
half-integer values for the ratio of the magneitc flux to the flux quantum
(for a similar structure of the persistent currents in carbon nanotube based
rings see, for example, Refs. \cite{Lin98}).

\begin{figure}[tbph]
\begin{center}
\begin{tabular}{cc}
\epsfig{figure=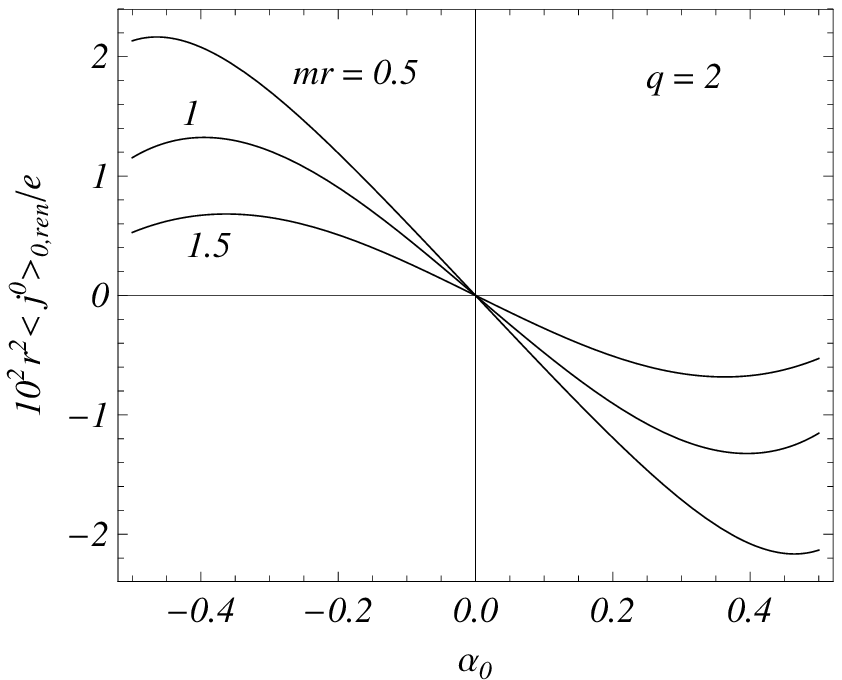,width=7.cm,height=6.cm} & \quad %
\epsfig{figure=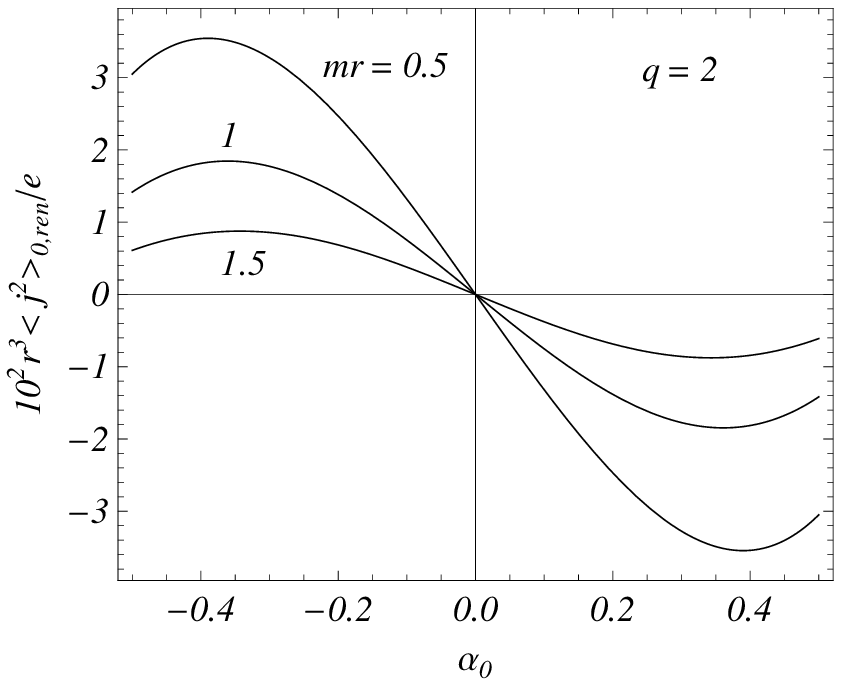,width=7.cm,height=6.cm}%
\end{tabular}%
\end{center}
\caption{The VEVs of the charge density (left panel) and the azimuthal
current (right panel) as functions of the magnetic flux for a massive
fermionic field in a boundary-free conical space with $q=2$.}
\label{fig3}
\end{figure}

Alternative expressions for the VEVs of the fermionic current are obtained
by using the integral representation (\ref{Rep2}) for the functions $%
\mathcal{I}(q,\pm \alpha _{0},z)$. We start with the charge density. The
corresponding regularized expression is given by Eq. (\ref{j00reg1}). As we
have already noticed, the first term on the right of this formula is finite
in the limit $s\rightarrow 0$. After the application of (\ref{Rep2}) to the
series in the second term, we see that the parts corresponding to the first
and last terms in the right-hand side of (\ref{Rep2}) vanish in the limit $%
s\rightarrow 0$ due to the exponential decay of the Macdonald functions. The
only term which survives in the limit $s\rightarrow 0$ is the part
corresponding to the second term in the right-hand side of Eq. (\ref{Rep2}).
In the expression for the fermionic current this term is multiplied by $q$
and hence it does not depend on the angle deficit and on the magnetic flux.
So, this term is the same as in the case of Minkowski spacetime in the
absence of the magnetic flux and is subtracted in the renormalization
procedure. As a result, for the renormalized charge density one finds the
expression below:%
\begin{eqnarray}
&&\langle j^{0}(x)\rangle _{0,\text{ren}}=-\frac{2em}{(2\pi )^{5/2}r}%
\int_{0}^{\infty }dz\,z^{-1/2}e^{-m^{2}r^{2}/2z-z}  \notag \\
&&\quad \times \Big[\text{sgn}(\alpha _{0})qB(q(|\alpha
_{0}|-1/2)+1/2,z)+2\int_{0}^{\infty }dy\,K_{iy}(z)g(q,\alpha _{0},y)\Big],
\label{j00renb}
\end{eqnarray}%
where
\begin{equation}
B(y,z)=\left\{
\begin{array}{cc}
0, & y\leqslant 0, \\
\sin (\pi y)K_{y}(z) & y>0,%
\end{array}%
\right.  \label{Byx}
\end{equation}%
and we have defined the function%
\begin{equation}
g(q,\alpha _{0},y)=\sum_{\delta =\pm 1}{\mathrm{Re}}\left[ \frac{\delta
\sinh (y\pi )}{e^{2\pi (y+i|q\delta \alpha _{0}-1/2|)/q}+1}\right] .
\label{gqalf}
\end{equation}

For the VEV of the azimuthal current, in a similar way, we get the
representation%
\begin{eqnarray}
&&\langle j^{2}(x)\rangle _{0,\text{ren}}=-\frac{4er^{-3}}{(2\pi )^{5/2}}%
\int_{0}^{\infty }dz\,z^{1/2}e^{-m^{2}r^{2}/2z-z}  \notag \\
&&\quad \times \Big[\text{sgn}(\alpha _{0})qB(q(|\alpha
_{0}|-1/2)+1/2,z)+2\int_{0}^{\infty }dy\,K_{iy}(z)g(q,\alpha _{0},y)\Big].
\label{j02renb}
\end{eqnarray}%
Note that under the condition (\ref{condalf0}) there are no square
integrable irregular modes and, in this case, the first terms in the square
brackets of Eqs. (\ref{j00renb}) and (\ref{j02renb}) vanish.

In the special case $q=1$ we see that $g(1,\alpha _{0},y)=0$, and for the
VEVs of the fermionic current we obtain the formulae%
\begin{eqnarray}
\langle j^{0}(x)\rangle _{0,\text{ren}} &=&-\frac{2em\sin (\pi \alpha _{0})}{%
(2\pi )^{5/2}r}\int_{0}^{\infty }dz\,e^{-m^{2}r^{2}/2z-z}\frac{K_{\alpha
_{0}}(z)}{\sqrt{z}},  \notag \\
\langle j^{2}(x)\rangle _{0,\text{ren}} &=&-\frac{4e\sin (\pi \alpha _{0})}{%
(2\pi )^{5/2}r^{3}}\int_{0}^{\infty }dz\,\sqrt{z}e^{-m^{2}r^{2}/2z-z}K_{%
\alpha _{0}}(z).  \label{j02renbm0}
\end{eqnarray}%
In the limit $\alpha _{0}\rightarrow \pm 1/2$ we recover results (\ref{Limj2}%
) and (\ref{Limj0}). In Appendix \ref{sec:App2New}, we show the equivalence
of these expressions to the ones previously given in the literature for the
fermionic densities induced by a magnetic flux in (2+1)-dimensional
Minkowski spacetime.

In the discussion above we used the irreducible representation of the
Clifford algebra corresponding to Eq. (\ref{DirMat}). For the second
representation the renormalized VEV of the azimuthal current is given by the
same expressions, whereas the expressions for the VEV of the renormalized
charge density change the sign. Consequently, the total induced charge (\ref%
{Q1}) changes the sign as well.

\section{Induced fermionic current in the exterior region}

\label{sec:ExtFC}

Now we turn to the investigation of the induced fermionic current in the
presence of a circular boundary at $r=a$ with boundary condition (\ref{BCMIT}%
). The corresponding VEV is evaluated using the mode sum formula (\ref%
{FCMode}) with the eigenspinors given by Eq. (\ref{psisig-}). For the
further discussion it is convenient to write these eigenspinors in an
equivalent form by using the properties (\ref{gsig1}):%
\begin{eqnarray}
\psi _{\gamma j}^{(+)}(x) &=&c_{0}e^{iqj\phi -iEt}\left(
\begin{array}{c}
g_{\beta _{j},\beta _{j}}(\gamma a,\gamma r)e^{-iq\phi /2} \\
\frac{\gamma \epsilon _{j}e^{iq\phi /2}}{E+m}g_{\beta _{j},\beta
_{j}+\epsilon _{j}}(\gamma a,\gamma r)%
\end{array}%
\right) ,  \notag \\
\psi _{\gamma j}^{(-)}(x) &=&c_{0}e^{-iqj\phi +iEt}\left(
\begin{array}{c}
\frac{\gamma \epsilon _{j}e^{-iq\phi /2}}{E+m}g_{\beta _{j},\beta
_{j}+\epsilon _{j}}(\gamma a,\gamma r) \\
g_{\beta _{j},\beta _{j}}(\gamma a,\gamma r)e^{iq\phi /2}%
\end{array}%
\right) ,  \label{psiplm}
\end{eqnarray}%
where $c_{0}$ is given by Eq. (\ref{c0}) with the replacement $|\lambda
_{n}|\rightarrow \beta _{j}$. In Eq. (\ref{psiplm}), $\epsilon _{j}$ is
defined by (\ref{epsj}) for $j\neq -\alpha $ and $\epsilon _{j}=-1$ for $%
j=-\alpha $. We could also obtain the representation (\ref{psiplm}), by
taking instead of Eq. (\ref{Zsig}) the linear combination of the functions $%
J_{\beta _{j}}(\gamma r)$ and $Y_{\beta _{j}}(\gamma r)$. Note that the
corresponding barred notation in expression (\ref{gsig}) may also be written
in the form%
\begin{equation}
\bar{F}_{\beta _{j}}^{(\pm )}(z)=-\epsilon _{j}zF_{\beta _{j}+\epsilon
_{j}}(z)\pm (\sqrt{z^{2}+\mu ^{2}}+\mu )F_{\beta _{j}}(z),  \label{barnot3}
\end{equation}%
with $F=J,Y$ and $\mu =ma$.

As in the boundary-free case, the VEV of the radial component vanishes and
for the charge density and the azimuthal current we have the expressions
\begin{eqnarray}
\langle j^{0}(x)\rangle &=&\frac{eq}{4\pi }\sum_{j}\int_{0}^{\infty }d\gamma
\frac{\gamma }{E}\frac{(E-m)g_{\beta _{j},\beta _{j}+\epsilon
_{j}}^{2}(\gamma a,\gamma r)+(E+m)g_{\beta _{j},\beta _{j}}^{2}(\gamma
a,\gamma r)}{\bar{J}_{\beta _{j}}^{(-)2}(\gamma a)+\bar{Y}_{\beta
_{j}}^{(-)2}(\gamma a)},  \notag \\
\langle j^{2}(x)\rangle &=&\frac{eq}{2\pi r}\sum_{j}\epsilon
_{j}\int_{0}^{\infty }d\gamma \frac{\gamma ^{2}}{E}\frac{g_{\beta _{j},\beta
_{j}}(\gamma a,\gamma r)g_{\beta _{j},\beta _{j}+\epsilon _{j}}(\gamma
a,\gamma r)}{\bar{J}_{\beta _{j}}^{(-)2}(\gamma a)+\bar{Y}_{\beta
_{j}}^{(-)2}(\gamma a)}.  \label{j02Ext}
\end{eqnarray}%
Here, as before, the summation goes over $j=\pm 1/2,\pm 3/2,\ldots $.
Equivalent forms are obtained using the eigenspinors (\ref{psisig-}). From
Eq. (\ref{j02Ext}) it follows that the fermionic current is a periodic
function of $\alpha $ with the period equal to 1. We assume that a cutoff
function is introduced without explicitly writing it. The specific form of
this function is not important for the discussion below.

In the presence of the boundary, the VEV of the fermionic current can be
decomposed as
\begin{equation}
\langle j^{\nu }(x)\rangle =\langle j^{\nu }(x)\rangle _{0}+\langle j^{\nu
}(x)\rangle _{\text{b}},  \label{jdecomp}
\end{equation}%
where $\langle j^{\nu }(x)\rangle _{\text{b}}$ is the part induced by the
boundary. In order to extract the latter explicitly, we subtract from Eq. (%
\ref{j02Ext}) the VEVs when the boundary is absent. If the ratio of the
magnetic flux to the flux quantum is not a half-integer, the boundary-free
parts are given by Eq. (\ref{j020}). In this case, for the evaluation of the
difference, in the expression of the charge density we use the identity%
\begin{equation}
\frac{g_{\beta _{j},\lambda }^{2}(x,y)}{\bar{J}_{\beta _{j}}^{(-)2}(x)+\bar{Y%
}_{\beta _{j}}^{(-)2}(x)}-J_{\lambda }^{2}(y)=-\frac{1}{2}\sum_{l=1,2}\frac{%
\bar{J}_{\beta _{j}}^{(-)}(x)}{\bar{H}_{\beta _{j}}^{(-,l)}(x)}H_{\lambda
}^{(l)2}(y),  \label{ident1}
\end{equation}%
with $\lambda =\beta _{j},\beta _{j}+\epsilon _{j}$, and with the Hankel
functions $H_{\nu }^{(l)}(x)$. The expression for the boundary-induced part
in the VEV of the charge density takes the form%
\begin{eqnarray}
&&\langle j^{0}(x)\rangle _{\text{b}}=-\frac{eq}{8\pi }\sum_{j}\sum_{l=1,2}%
\int_{0}^{\infty }d\gamma \frac{\gamma }{E}\frac{\bar{J}_{\beta
_{j}}^{(-)}(\gamma a)}{\bar{H}_{\beta _{j}}^{(-,l)}(\gamma a)}  \notag \\
&&\qquad \times \left[ (E-m)H_{\beta _{j}+\epsilon _{j}}^{(l)2}(\gamma
r)+(E+m)H_{\beta _{j}}^{(l)2}(\gamma r)\right] .  \label{j001}
\end{eqnarray}%
Now, in the complex plane $\gamma $, we rotate the integration contour by
the angle $\pi /2$ for the term with $l=1$ and by the angle $-\pi /2$ for
the term with $l=2$. The integrals over the segments $(0,im)$ and $(0,-im)$
cancel each other and, introducing the modified Bessel functions, we get the
following expression%
\begin{eqnarray}
&&\langle j^{0}(x)\rangle _{\text{b}}=-\frac{eq}{2\pi ^{2}}%
\sum_{j}\int_{m}^{\infty }dz\,z  \notag \\
&&\quad \times \Big\{m\frac{K_{\beta _{j}}^{2}(zr)+K_{\beta _{j}+\epsilon
_{j}}^{2}(zr)}{\sqrt{z^{2}-m^{2}}}{\mathrm{Re}}\left[ I_{\beta
_{j}}^{(-)}(za)/K_{\beta _{j}}^{(-)}(za)\right]  \notag \\
&&\quad -\left[ K_{\beta _{j}}^{2}(zr)-K_{\beta _{j}+\epsilon _{j}}^{2}(zr)%
\right] {\mathrm{Im}}\left[ I_{\beta _{j}}^{(-)}(za)/K_{\beta _{j}}^{(-)}(za)%
\right] \Big\},  \label{j002b}
\end{eqnarray}%
where we use the notation (the notation with the upper sign is used in the
next section)%
\begin{equation}
F^{(\pm )}(z)=zF^{\prime }(z)+\left( \pm \mu \pm i\sqrt{z^{2}-\mu ^{2}}%
-\epsilon _{j}\beta _{j}\right) F(z).  \label{F+}
\end{equation}

The ratio in the integrand of Eq. (\ref{j002b}) can also be written in the
form%
\begin{equation}
\frac{I_{\beta _{j}}^{(-)}(x)}{K_{\beta _{j}}^{(-)}(x)}=\frac{W_{\beta
_{j},\beta _{j}+\epsilon _{j}}^{(-)}(x)+i\sqrt{1-\mu ^{2}/x^{2}}}{x[K_{\beta
_{j}}^{2}(x)+K_{\beta _{j}+\epsilon _{j}}^{2}(x)]+2\mu K_{\beta
_{j}}(x)K_{\beta _{j}+\epsilon _{j}}(x)},  \label{IKratio}
\end{equation}%
with the notation%
\begin{eqnarray}
W_{\beta _{j},\beta _{j}+\epsilon _{j}}^{(\pm )}(x) &=&x\left[ I_{\beta
_{j}}(x)K_{\beta _{j}}(x)-I_{\beta _{j}+\epsilon _{j}}(x)K_{\beta
_{j}+\epsilon _{j}}(x)\right]  \notag \\
&&\pm \mu \left[ I_{\beta _{j}+\epsilon _{j}}(x)K_{\beta _{j}}(x)-I_{\beta
_{j}}(x)K_{\beta _{j}+\epsilon _{j}}(x)\right] .  \label{Wbet}
\end{eqnarray}%
The real and imaginary parts appearing in Eq. (\ref{j002b}) are easily
obtained from Eq. (\ref{IKratio}). Note that under the change $\alpha
\rightarrow -\alpha $, $j\rightarrow -j$, we have $\beta _{j}\rightarrow
\beta _{j}+\epsilon _{j}$, $\beta _{j}+\epsilon _{j}\rightarrow \beta _{j}$.
From here it follows that the real/imaginary part in Eq. (\ref{IKratio}) is
an odd/even function under this change. Now, from Eq. (\ref{j002b}) we see
that the boundary-induced part in the VEV is an odd function of $\alpha $.
When $\alpha $ is a half-integer, in the term of Eq. (\ref{j002b}) with $%
j=-\alpha $ the orders of the modified Bessel functions are equal to $\pm
1/2 $. Using the expressions of these functions in terms of the elementary
functions, we can see that the corresponding integral vanishes. For the
terms with $j\neq -\alpha $, the contributions of $j<-\alpha $ and $%
j>-\alpha $ to the right-hand side of Eq. (\ref{j002b}) cancel each other.
Hence, if $\alpha $ is a half-integer the boundary-induced part in the VEV
of the charge density vanishes. Recall that the same is the case for the
boundary-free part.

In a similar way, using the identity%
\begin{equation}
\frac{g_{\beta _{j}}(x,y)g_{\beta _{j}+\epsilon _{j}}(x,y)}{\bar{J}_{\beta
_{j}}^{(-)2}(x)+\bar{Y}_{\beta _{j}}^{(-)2}(x)}=J_{\beta _{j}}(y)J_{\beta
_{j}+\epsilon _{j}}(y)-\frac{1}{2}\sum_{l=1,2}\frac{\bar{J}_{\beta
_{j}}^{(-)}(x)}{\bar{H}_{\beta _{j}}^{(-,l)}(x)}H_{\beta
_{j}}^{(l)}(y)H_{\beta _{j}+\epsilon _{j}}^{(l)}(y),  \label{ident2}
\end{equation}%
for the boundary-induced part in the VEV of the azimuthal current we find%
\begin{eqnarray}
&&\langle j^{2}(x)\rangle _{\text{b}}=-\frac{eq}{\pi ^{2}r}%
\sum_{j}\int_{m}^{\infty }dz\frac{z^{2}}{\sqrt{z^{2}-m^{2}}}  \notag \\
&&\qquad \times K_{\beta _{j}}(zr)K_{\beta _{j}+\epsilon _{j}}(zr){\mathrm{Re%
}}\left[ I_{\beta _{j}}^{(-)}(za)/K_{\beta _{j}}^{(-)}(za)\right] .
\label{j21}
\end{eqnarray}%
As in the case of charge density, this part is a periodical function of the
magnetic flux with a period equal the flux quantum.

If we present the ratio of the magnetic flux to the flux quantum in the form
(\ref{alf0}), then the boundary induced VEVs are functions of $\alpha _{0}$
alone. They are odd functions of this parameter. In the limit $\alpha
_{0}\rightarrow \pm 1/2$, $|\alpha _{0}|<1/2$, the only nonzero contribution
to $\langle j^{\nu }(x)\rangle _{\text{b}}$ comes from the term with $j=\mp
1/2$ and one has the limiting values%
\begin{eqnarray}
\lim_{\alpha _{0}\rightarrow \pm 1/2}\langle j^{0}(x)\rangle _{\text{b}}
&=&\pm \frac{eqm}{2\pi ^{2}r}K_{0}(2mr),  \notag \\
\lim_{\alpha _{0}\rightarrow \pm 1/2}\langle j^{2}(x)\rangle _{\text{b}}
&=&\pm \frac{eqm}{2\pi ^{2}r^{2}}K_{1}(2mr).  \label{Limj02b}
\end{eqnarray}%
Now comparing with Eqs. (\ref{Limj2}) and (\ref{Limj0}), we see that the
limiting value of the total current density (\ref{jdecomp}) is zero and the
latter is continuous at $\alpha _{0}=\pm 1/2$. This result was expected due
to the continuity of the exterior eigenspinors as functions of the parameter
$\alpha _{0}$. Comparing with the results of the previous section, we see
that the limiting transitions $a\rightarrow 0$ and $|\alpha _{0}|\rightarrow
1/2$ do not commute.

For a massless field the expressions for the boundary-induced parts in the
VEVs take the form%
\begin{eqnarray}
\langle j^{0}(x)\rangle _{\text{b}} &=&\frac{eq}{2\pi ^{2}a^{2}}%
\sum_{j}\int_{0}^{\infty }dz\,\frac{K_{\beta _{j}}^{2}(zr/a)-K_{\beta
_{j}+\epsilon _{j}}^{2}(zr/a)}{K_{\beta _{j}}^{2}(z)+K_{\beta _{j}+\epsilon
_{j}}^{2}(z)},  \notag \\
\langle j^{2}(x)\rangle _{\text{b}} &=&-\frac{eq}{\pi ^{2}a^{2}r}%
\sum_{j}\int_{0}^{\infty }dz\,\frac{K_{\beta _{j}}(zr/a)K_{\beta
_{j}+\epsilon _{j}}(zr/a)}{K_{\beta _{j}}^{2}(z)+K_{\beta _{j}+\epsilon
_{j}}^{2}(z)}W_{\beta _{j},\beta _{j}+\epsilon _{j}}^{(-)}(z),
\label{J02Extm0}
\end{eqnarray}%
with the notation defined by Eq. (\ref{Wbet}). We would like to point out
that the boundary-induced charge density does not vanish for a massless
filed. The corresponding boundary-free part vanishes and, hence, $\langle
j^{0}(x)\rangle =\langle j^{0}(x)\rangle _{\text{b}}$.

Now we turn to the investigation of the boundary-induced part in the VEV of
fermionic current in the asymptotic regions of the parameters. In the limit $%
a\rightarrow 0$, for fixed values of $r$, by taking into account that
\begin{equation}
\frac{I_{\beta _{j}}^{(-)}(za)}{K_{\beta _{j}}^{(-)}(za)}\approx a\frac{%
\epsilon _{j}m+i\sqrt{z^{2}-m^{2}}}{\Gamma ^{2}(q|j+\alpha |+1/2)}%
(za/2)^{2q|j+\alpha |-1},  \label{IKpla0}
\end{equation}%
to the leading order, from Eqs. (\ref{j002b}) and (\ref{j21}), we have%
\begin{eqnarray}
\langle j^{0}(x)\rangle _{\text{b}} &\approx &\frac{eq}{\pi ^{2}}\frac{\text{%
sgn}(\alpha _{0})(a/2r)^{2q_{\alpha }}}{r^{2}\Gamma ^{2}(q_{\alpha }+1/2)}%
\int_{mr}^{\infty }dz\,\frac{z^{2q_{\alpha }}}{\sqrt{z^{2}-m^{2}r^{2}}}
\notag \\
&&\times \left[ \left( 2m^{2}r^{2}-z^{2}\right) K_{q_{\alpha
}-1/2}^{2}(z)+z^{2}K_{q_{\alpha }+1/2}^{2}(z)\right] ,  \label{j0bExta0} \\
\langle j^{2}(x)\rangle _{\text{b}} &\approx &\frac{2eqm}{\pi ^{2}r^{2}}%
\frac{\text{sgn}(\alpha _{0})(a/2r)^{2q_{\alpha }}}{\Gamma ^{2}(q_{\alpha
}+1/2)}\int_{mr}^{\infty }dz\frac{z^{2q_{\alpha }+1}}{\sqrt{z^{2}-m^{2}r^{2}}%
}K_{q_{\alpha }-1/2}(z)K_{q_{\alpha }-1/2}(z),  \notag
\end{eqnarray}%
with the notation%
\begin{equation}
q_{\alpha }=q(1/2-|\alpha _{0}|).  \label{qalfa}
\end{equation}%
For a massless field the asymptotic behavior for the charge density is
directly obtained from Eq. (\ref{j0bExta0}). The integrals involving the
Macdonald function are evaluated in terms of the gamma function and one finds%
\begin{equation}
\langle j^{0}(x)\rangle _{\text{b}}\approx \frac{eq\,\text{sgn}(\alpha _{0})%
}{\pi r^{2}}\left( \frac{a}{2r}\right) ^{2q_{\alpha }}\frac{q_{\alpha
}\Gamma (2q_{\alpha }+1/2)\Gamma (q_{\alpha }+1)}{(2q_{\alpha }+1)\Gamma
^{3}(q_{\alpha }+1/2)}.  \label{j0bExta0m0}
\end{equation}%
For the azimuthal component the leading term in Eq. (\ref{j0bExta0})
vanishes. The corresponding asymptotic behavior is directly found from Eq. (%
\ref{J02Extm0}). The leading term is given by
\begin{equation}
\langle j^{2}(x)\rangle _{\text{b}}\approx \frac{2eq}{\pi r^{3}}\frac{\text{%
sgn}(\alpha _{0})}{(2q_{\alpha })^{2}-1}\left( \frac{a}{2r}\right)
^{2q_{\alpha }+1}\frac{\Gamma (2q_{\alpha }+3/2)\Gamma (q_{\alpha }+1)}{%
(2q_{\alpha }+1)\Gamma ^{3}(q_{\alpha }+1/2)},  \label{j2bExtm0}
\end{equation}%
for $q_{\alpha }>1/2$, and by the expression%
\begin{equation}
\langle j^{2}(x)\rangle _{\text{b}}\approx -\frac{eq\,\text{sgn}(\alpha _{0})%
}{2^{2q_{\alpha }+1}\pi r^{3}}\left( \frac{a}{2r}\right) ^{4q_{\alpha }}%
\frac{\Gamma (1/2-q_{\alpha })}{\Gamma ^{4}(1/2+q_{\alpha })}\Gamma
(2q_{\alpha }+1/2)\Gamma (3q_{\alpha }+1),  \label{j2bExtm0b}
\end{equation}%
in the case $q_{\alpha }<1/2$. For for $q_{\alpha }=1/2$, the leading terms
behaves as $a^{2}\ln (a)$.

\begin{figure}[tbph]
\begin{center}
\begin{tabular}{cc}
\epsfig{figure=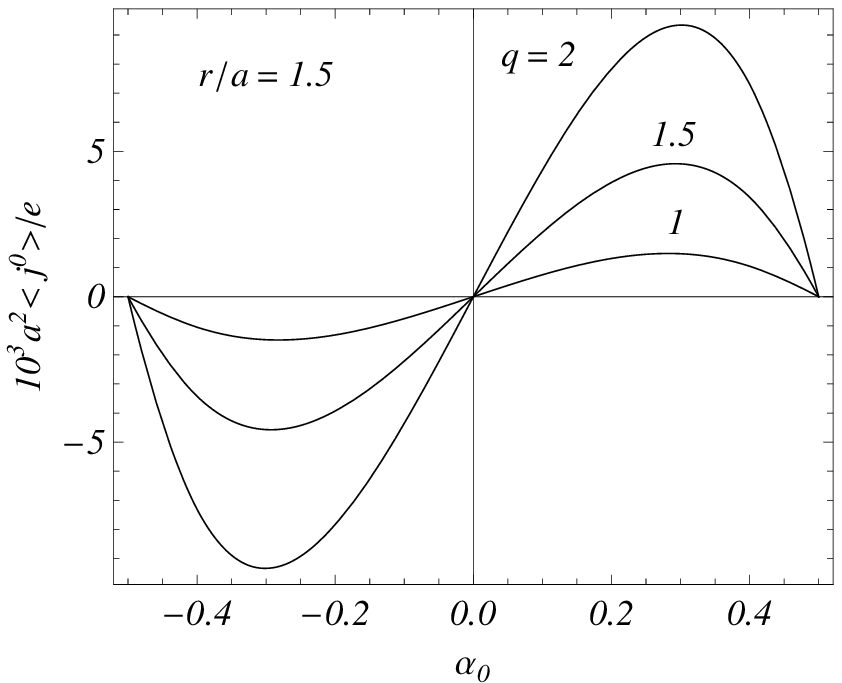,width=7.cm,height=6.cm} & \quad %
\epsfig{figure=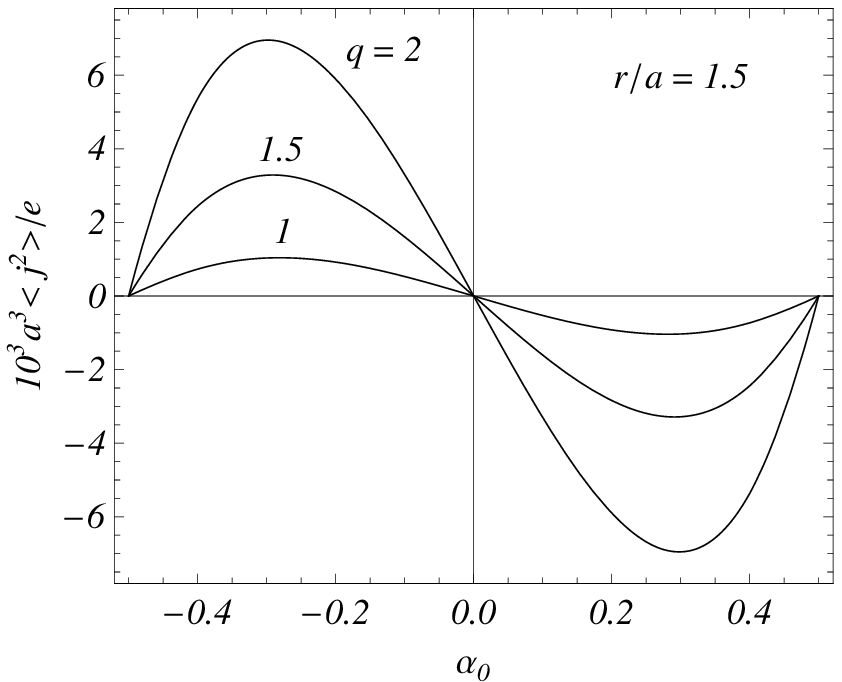,width=7.cm,height=6.cm}%
\end{tabular}%
\end{center}
\caption{The VEVs of the charge density (left panel) and azimuthal current
(right panel) for a massless fermionic field outside a circular boundary.}
\label{fig4}
\end{figure}

At large distances from the boundary, for a massive field, under the
condition $mr\gg 1$, the dominant contribution to the integrals come from
the region near the lower limit of the integration and to the leading order
we find%
\begin{eqnarray}
\langle j^{0}(x)\rangle _{\text{b}} &\approx &-\frac{eqm^{2}e^{-2rm}}{4\sqrt{%
\pi }(rm)^{3/2}}\sum_{j}{\mathrm{Re}}\left[ I_{\beta
_{j}}^{(-)}(ma)/K_{\beta _{j}}^{(-)}(ma)\right] ,  \notag \\
\langle j^{2}(x)\rangle _{\text{b}} &\approx &-\frac{eqm^{3}e^{-2rm}}{4\sqrt{%
\pi }(mr)^{5/2}}\sum_{j}{\mathrm{Re}}\left[ I_{\beta
_{j}}^{(-)}(ma)/K_{\beta _{j}}^{(-)}(ma)\right] .  \label{j02LargeDist}
\end{eqnarray}%
As expected, we have an exponential suppression of the boundary-induced
VEVs. For a massless field, the asymptotics at large distances are given by
Eqs. (\ref{j0bExta0m0})-(\ref{j2bExtm0b}). In Fig. \ref{fig4}, we plot the
VEVs of the charge density (left panel) and azimuthal current (right panel)
for a massless fermionic field as functions of the magnetic flux. The graphs
are plotted for $r/a=1.5$ and for several values of the parameter $q$. As we
have already mentioned, in the exterior region the total VEVs of the charge
density and azimuthal current vanish in the limt $|\alpha _{0}|\rightarrow 0$%
. Note that for a massless field the boundary-free part in the VEV\ of the
charge density vanishes and the non-zero charge density on the left plot is
induced by the circular boundary.

\section{Fermionic current inside a circular boundary}

\label{sec:Int}

In this section we consider the region inside a circular boundary with
radius $a$, $r<a$, on which the fermionic field obeys the boundary condition
(\ref{BCMIT}) with $n_{\mu }=-\delta _{\mu }^{1}$. The boundary condition at
the cone apex for the irregular mode is the same as that we have used in
Section \ref{sec:BoundFree} for the boundary-free conical geometry. The
eigenspinors in this region have the form%
\begin{eqnarray}
\psi _{\gamma j}^{(+)} &=&\varphi _{0}e^{iqj\phi -iEt}\left(
\begin{array}{c}
e^{-iq\phi /2}J_{\beta _{j}}(\gamma r) \\
\frac{\gamma \epsilon _{j}e^{iq\phi /2}}{E+m}J_{\beta _{j}+\epsilon
_{j}}(\gamma r)%
\end{array}%
\right) ,  \notag \\
\psi _{\gamma j}^{(-)} &=&\varphi _{0}e^{-iqj\phi +iEt}\left(
\begin{array}{c}
\frac{\epsilon _{j}\gamma e^{-iq\phi /2}}{E+m}J_{\beta _{j}+\epsilon
_{j}}(\gamma r) \\
e^{iq\phi /2}J_{\beta _{j}}(\gamma r)%
\end{array}%
\right) ,  \label{psiInt}
\end{eqnarray}%
with the same notations as in Section \ref{sec:BoundFree}. From the boundary
condition at $r=a$ we find that the eigenvalues of $\gamma $ are solutions
of the equation%
\begin{equation}
J_{\beta _{j}}(\gamma a)-\frac{\gamma \epsilon _{j}}{E+m}J_{\beta
_{j}+\epsilon _{j}}(\gamma a)=0.  \label{gamVal}
\end{equation}%
Note that this equation may also be written in the form $\bar{J}_{\beta
_{j}}^{(+)}(\gamma a)=0$ with the barred notation defined by Eq. (\ref%
{barnot2}). For a given $\beta _{j}$, Eq. (\ref{gamVal}) has an infinite
number of solutions which we denote by $\gamma a=\gamma _{\beta _{j},l}$, $%
l=1,2,\ldots $.

The normalization coefficient in Eq. (\ref{psiInt}) is determined from the
condition
\begin{equation}
\int_{0}^{a}dr\int_{0}^{\phi _{0}}d\phi \,r\psi _{\gamma j}^{(\pm )\dagger
}\psi _{\gamma ^{\prime }j^{\prime }}^{(\pm )}=\delta _{ll^{\prime }}\delta
_{jj^{\prime }}\ .  \label{NormInt}
\end{equation}%
Using the standard integral for the square of the Bessel function (see, for
example, \cite{Prud86}), one finds%
\begin{equation}
\varphi _{0}^{2}=\frac{a^{2}y^{2}}{\phi _{0}J_{\beta _{j}}^{2}(y)}\left[
2(y^{2}+\mu ^{2})-\left( 2\epsilon _{j}\beta _{j}+1\right) \sqrt{y^{2}+\mu
^{2}}+\mu \right] ^{-1},\;y=\gamma _{\beta _{j},l},  \label{NormCoefInt}
\end{equation}%
where, as before, $\mu =ma$. For the further convenience we will write this
expression in the form%
\begin{equation}
\varphi _{0}^{2}=\frac{yT_{\beta _{j}}(y)}{2\phi _{0}a^{2}}\frac{\mu +\sqrt{%
y^{2}+\mu ^{2}}}{\sqrt{y^{2}+\mu ^{2}}},  \label{phi0T}
\end{equation}%
with the notation%
\begin{equation}
T_{\beta _{j}}(y)=\frac{y}{J_{\beta _{j}}^{2}(y)}\left[ y^{2}+\left( \mu
-\epsilon _{j}\beta _{j}\right) \left( \mu +\sqrt{y^{2}+\mu ^{2}}\right) -%
\frac{y^{2}}{2\sqrt{y^{2}+\mu ^{2}}}\right] ^{-1}.  \label{Tnu}
\end{equation}

Substituting the eigenspinors (\ref{psiInt}) into the mode sum formula%
\begin{equation}
\langle j^{\nu }(x)\rangle =e\sum_{j}\sum_{l=1}^{\infty }\,\bar{\psi}%
_{\gamma j}^{(-)}(x)\gamma ^{\nu }\psi _{\gamma j}^{(-)}(x),
\label{modesumInt}
\end{equation}%
for the VEVs of separate components of fermionic current we have%
\begin{eqnarray}
\langle j^{0}(x)\rangle &=&\frac{eq}{4\pi a^{2}}\sum_{j}\sum_{l=1}^{\infty
}yT_{\beta _{j}}(y)\Big[\Big(\frac{\mu }{\sqrt{y^{2}+\mu ^{2}}}+1\Big)%
J_{\beta _{j}}^{2}(yr/a)-\Big(\frac{\mu }{\sqrt{y^{2}+\mu ^{2}}}-1\Big)%
J_{\beta _{j}+\epsilon _{j}}^{2}(yr/a)\Big],  \notag \\
\langle j^{2}(x)\rangle &=&\frac{eq}{2\pi a^{2}r}\sum_{j}\sum_{l=1}^{\infty }%
\frac{\epsilon _{j}y^{2}T_{\beta _{j}}(y)}{\sqrt{y^{2}+\mu ^{2}}}J_{\beta
_{j}}(yr/a)J_{\beta _{j}+\epsilon _{j}}(yr/a),  \label{j2Int}
\end{eqnarray}%
with $y=\gamma _{\beta _{j},l}$, and the radial component vanishes, $\langle
j^{1}(x)\rangle =0$. As before, we assume the presence of a cutoff function
without explicitly writing it.

As the explicit form for $\gamma _{\beta _{j},l}$ is not known, Eqs. (\ref%
{j2Int}) are not convenient for the direct evaluation of the VEVs. In
addition, the separate terms in the mode sum are highly oscillatory for
large values of the quantum numbers. In order to find a convenient integral
representation, we apply to the series over $l$ the summation formula (see
\cite{Saha04,Saha08Book})%
\begin{eqnarray}
&&\sum_{l=1}^{\infty }f(\gamma _{\beta _{j},l})T_{\beta }(\gamma _{\beta
_{j},l})=\int_{0}^{\infty }dx\,f(x)-\frac{1}{\pi }\int_{0}^{\infty }dx
\notag \\
&&\quad \times \bigg[e^{-\beta _{j}\pi i}f(xe^{\pi i/2})\frac{K_{\beta
_{j}}^{(+)}(x)}{I_{\beta _{j}}^{(+)}(x)}+e^{\beta _{j}\pi i}f(xe^{-\pi i/2})%
\frac{K_{\beta _{j}}^{(+)\ast }(x)}{I_{\beta _{j}}^{(+)\ast }(x)}\bigg],
\label{SumForm}
\end{eqnarray}%
the asterisk meaning complex conjugate. Here the notation $F^{(+)}(x)$ for a
given function $F(x)$ is defined by Eq. (\ref{F+}) for $x\geqslant \mu $ and
by the relation
\begin{equation}
F^{(+)}(x)=xF^{\prime }(x)+(\mu +\sqrt{\mu ^{2}-x^{2}}-\epsilon _{j}\beta
_{j})F(x),  \label{F+2}
\end{equation}%
for $x<\mu $. Note that in the latter case $F^{(+)\ast }(x)=F^{(+)}(x)$. The
term in the VEVs corresponding to the first integral in the right-hand side
of Eq. (\ref{SumForm}) coincides with the VEV of the fermionic current for
the situation where the boundary is absent.

As a result, the VEV of the fermionic current is presented in the decomposed
form (\ref{jdecomp}). For the function $f(x)$ corresponding to Eq. (\ref%
{j2Int}), in the second term on the right-hand side of Eq. (\ref{SumForm}),
the part of the integral over the region $(0,\mu )$ vanishes. Consequently,
the boundary-induced contribution for the charge density in the region
inside the circle is given by the expression
\begin{eqnarray}
&&\langle j^{0}(x)\rangle _{\text{b}}=-\frac{eq}{2\pi ^{2}}%
\sum_{j}\int_{m}^{\infty }dz\,z  \notag \\
&&\qquad \times \Big\{m\frac{I_{\beta _{j}}^{2}(zr)+I_{\beta _{j}+\epsilon
_{j}}^{2}(zr)}{\sqrt{z^{2}-m^{2}}}{\mathrm{Re}}[K_{\beta
_{j}}^{(+)}(za)/I_{\beta _{j}}^{(+)}(za)]  \notag \\
&&\qquad -[I_{\beta _{j}}^{2}(zr)-I_{\beta _{j}+\epsilon _{j}}^{2}(zr)]{%
\mathrm{Im}}[K_{\beta _{j}}^{(+)}(za)/I_{\beta _{j}}^{(+)}(za)]\Big\}.
\label{j0int1}
\end{eqnarray}%
Similarly, for the boundary-induced part in the azimuthal component we find%
\begin{equation}
\langle j^{2}(x)\rangle _{\text{b}}=\frac{eq}{\pi ^{2}r}\sum_{j}\int_{m}^{%
\infty }dz\,\frac{z^{2}I_{\beta _{j}}(zr)I_{\beta _{j}+\epsilon _{j}}(zr)}{%
\sqrt{z^{2}-m^{2}}}{\mathrm{Re}}[K_{\beta _{j}}^{(+)}(za)/I_{\beta
_{j}}^{(+)}(za)].  \label{j2int1}
\end{equation}%
For points away from the circular boundary, the boundary-induced
contributions (\ref{j0int1}) and (\ref{j2int1}) are finite and the
renormalization is reduced to that for the boundary-free geometry. These
contributions are periodic functions of the parameter $\alpha $ with the
period equal to 1. So, if we present this parameter in the form (\ref{alf0})
with $n_{0}$ being an integer, then the VEVs depend on $\alpha _{0}$ alone
and they are odd functions of this parameter. Similar to Eq. (\ref{IKratio}%
), the ratio of the combinations of the modified Bessel functions in Eq. (%
\ref{j0int1}) is presented in the form%
\begin{equation}
\frac{K_{\beta _{j}}^{(+)}(x)}{I_{\beta _{j}}^{(+)}(x)}=\frac{W_{\beta
_{j},\beta _{j}+\epsilon _{j}}^{(+)}(x)+i\sqrt{1-\mu ^{2}/x^{2}}}{x[I_{\beta
_{j}}^{2}(x)+I_{\beta _{j}+\epsilon _{j}}^{2}(x)]+2\mu I_{\beta
_{j}}(x)I_{\beta _{j}+\epsilon _{j}}(x)},  \label{KIratio}
\end{equation}%
with the notation defined by Eq. (\ref{Wbet}).

When the parameter $\alpha $ is a half-integer the contributions of the
modes with $j\neq -\alpha $ to the boundary-induced VEVs inside the circle
are still given by expressions (\ref{j0int1}) and (\ref{j2int1}). It can be
easily seen that the contributions of the modes with $j<-\alpha $ and $%
j>-\alpha $ cancel each other. The contribution of the mode with $j=-\alpha $
should be considered separately. In Appendix \ref{sec:App2} we show that
this contribution vanishes as well. Therefore, we conclude that for $\alpha $
being a half-integer, the boundary-induced part in the VEV of the fermionic
current vanishes.

For a massless field the formulae of the boundary-induced parts are
simplified to%
\begin{eqnarray}
\langle j^{0}(x)\rangle _{\text{b}} &=&\frac{eq}{2\pi ^{2}a^{2}}%
\sum_{j}\int_{0}^{\infty }dz\,\frac{I_{\beta _{j}}^{2}(zr/a)-I_{\beta
_{j}+\epsilon _{j}}^{2}(zr/a)}{I_{\beta _{j}}^{2}(z)+I_{\beta _{j}+\epsilon
_{j}}^{2}(z)},  \notag \\
\langle j^{2}(x)\rangle _{\text{b}} &=&\frac{eq}{\pi ^{2}ra^{2}}%
\sum_{j}\int_{0}^{\infty }dz\,\frac{I_{\beta _{j}}(zr/a)I_{\beta
_{j}+\epsilon _{j}}(zr/a)}{I_{\beta _{j}}^{2}(z)+I_{\beta _{j}+\epsilon
_{j}}^{2}(z)}W_{\beta _{j},\beta _{j}+\epsilon _{j}}^{(+)}(z).
\label{j02intm0}
\end{eqnarray}%
Note that for a massless field $W_{\beta _{j},\beta _{j}+\epsilon
_{j}}^{(+)}(z)=W_{\beta _{j},\beta _{j}+\epsilon _{j}}^{(-)}(z)$. In the
limit $\alpha _{0}\rightarrow \pm 1/2$, $|\alpha _{0}|<1/2$, the only
nonzero contributions to Eqs. (\ref{j0int1}) and (\ref{j2int1}) come from
the term with $j=\mp 1/2$ and, by making use of Eqs. (\ref{Limj2}), for the
total VEVs we find%
\begin{eqnarray}
\lim_{\alpha _{0}\rightarrow \pm 1/2}\langle j^{0}(x)\rangle &=&\mp \frac{eqm%
}{2\pi ^{2}r}K_{0}(2mr)\pm \frac{eq}{\pi ^{2}r}\int_{m}^{\infty }dz\,\frac{%
amz\cosh (2zr)-(z+m)e^{2za}}{\sqrt{z^{2}-m^{2}}(\frac{z+m}{z-m}e^{4za}+1)},
\notag \\
\lim_{\alpha _{0}\rightarrow \pm 1/2}\langle j^{2}(x)\rangle &=&\mp \frac{eqm%
}{2\pi ^{2}r^{2}}K_{1}(2mr)\mp \frac{eqa}{\pi ^{2}r^{2}}\int_{m}^{\infty
}dz\,\frac{z^{2}}{\sqrt{z^{2}-m^{2}}}\frac{\sinh (2zr)}{\frac{z+m}{z-m}%
e^{4za}+1}.  \label{j02intLim}
\end{eqnarray}%
For a massless field they reduce to the expressions:%
\begin{eqnarray}
\lim_{\alpha _{0}\rightarrow \pm 1/2}\langle j^{0}(x)\rangle &=&\mp \frac{eq%
}{8a\pi r},  \notag \\
\lim_{\alpha _{0}\rightarrow \pm 1/2}\langle j^{2}(x)\rangle &=&\mp \frac{eq%
}{8\pi ^{2}r^{3}}\left[ 1+\frac{\pi r/2a}{\sin (\pi r/2a)}\right] .
\label{j02intLim0}
\end{eqnarray}%
Note that the limiting values are linear functions of the parameter $q$.

The general expressions for the VEVs are simplified in asymptotic regions of
the parameters. First we consider large values of the circle radius. For the
modified Bessel functions in the integrands of Eqs. (\ref{j0int1}) and (\ref%
{j2int1}), with $za$ in their arguments, we use the asymptotic expansions
for large values of the argument. By taking into account that for a massive
field the dominant contribution into the integrals comes from the
integration range near the lower limit, to the leading order we find%
\begin{eqnarray}
\langle j^{0}(x)\rangle _{\text{b}} &\approx &\frac{eqm^{2}e^{-2ma}}{8\sqrt{%
\pi }(ma)^{3/2}}\sum_{j}\epsilon _{j}\left[ (\beta _{j}+\epsilon
_{j})I_{\beta _{j}}^{2}(mr)+\beta _{j}I_{\beta _{j}+\epsilon _{j}}^{2}(mr)%
\right] ,  \notag \\
\langle j^{2}(x)\rangle _{\text{b}} &\approx &-\frac{eqm^{2}e^{-2ma}}{8\sqrt{%
\pi }r(ma)^{3/2}}\sum_{j}\,(2\epsilon _{j}\beta _{j}+1)I_{\beta
_{j}}(mr)I_{\beta _{j}+\epsilon _{j}}(mr).  \label{j02LargeRad}
\end{eqnarray}%
In this limit, for a fixed value of the radial coordinate, the
boundary-induced VEVs decay exponentially. For a massless field, assuming $%
r/a\ll 1$, we expand the modified Bessel function in the numerators of
integrands in Eq. (\ref{j02intm0}) in powers of $r/a$. The dominant
contribution comes from the term $j=1/2$ for $\alpha _{0}<0$ and from the
term $j=-1/2$ for $\alpha _{0}>0$. To the leading order we have%
\begin{eqnarray}
\langle j^{0}(x)\rangle _{\text{b}} &\approx &-\frac{eq}{2\pi ^{2}a^{2}}%
\frac{\text{sgn}(\alpha _{0})(r/2a)^{2q_{\alpha }-1}}{\Gamma ^{2}(q_{\alpha
}+1/2)}\int_{0}^{\infty }dz\,\frac{z^{2q_{\alpha }-1}}{I_{q_{\alpha
}+1/2}^{2}(z)+I_{q_{\alpha }-1/2}^{2}(z)},  \notag \\
\langle j^{2}(x)\rangle _{\text{b}} &\approx &-\frac{eq}{\pi ^{2}a^{3}}\frac{%
\text{sgn}(\alpha _{0})(r/2a)^{2q_{\alpha }-1}}{(2q_{\alpha }+1)\Gamma
^{2}(q_{\alpha }+1/2)}\int_{0}^{\infty }dz\,\frac{z^{2q_{\alpha
}}W_{q_{\alpha }-1/2,q_{\alpha }+1/2}^{(+)}(z)}{I_{q_{\alpha
}+1/2}^{2}(z)+I_{q_{\alpha }-1/2}^{2}(z)},  \label{j02LargeRadm0}
\end{eqnarray}%
where $q_{\alpha }$ is defined in Eq. (\ref{qalfa}). As it is seen, for a
massless field the decay of the VEVs is as power-law.

\begin{figure}[tbph]
\begin{center}
\begin{tabular}{cc}
\epsfig{figure=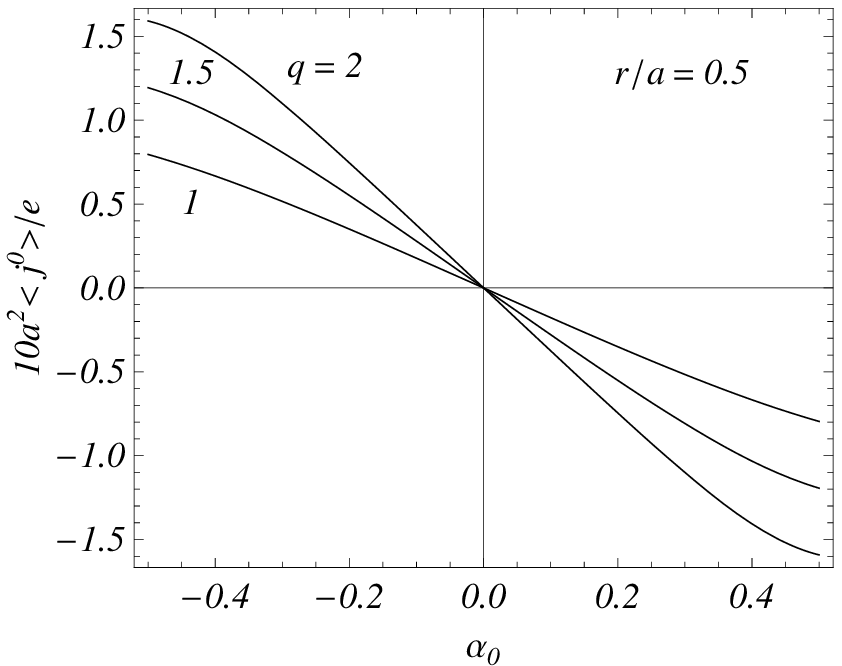,width=7.cm,height=6.cm} & \quad %
\epsfig{figure=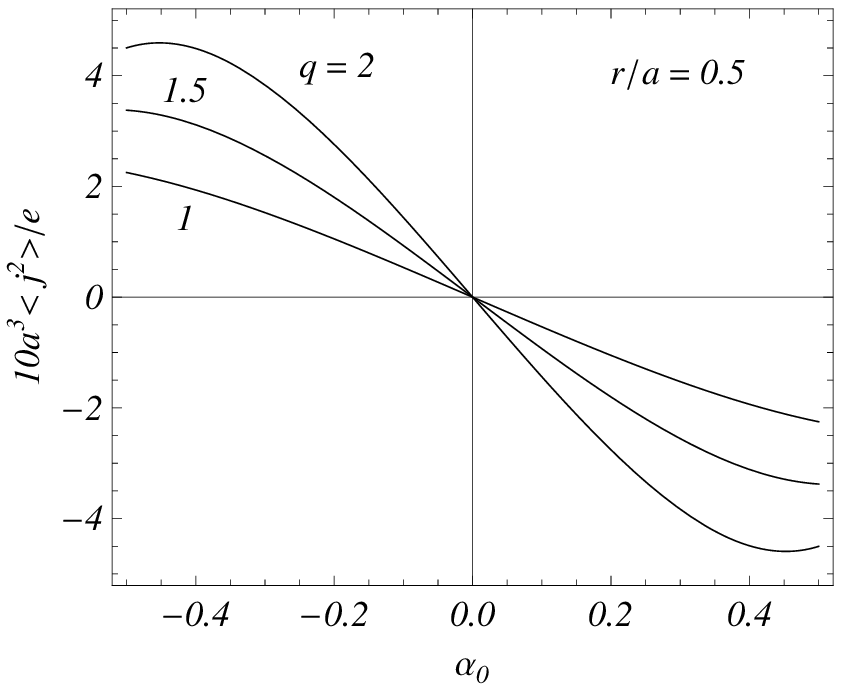,width=7.cm,height=6.cm}%
\end{tabular}%
\end{center}
\caption{The same as in figure \protect\ref{fig4} for the region inside a
circular boundary. }
\label{fig5}
\end{figure}

For points near the apex of the cone, $r\rightarrow 0$, we have the
following leading terms%
\begin{eqnarray}
&&\langle j^{0}(x)\rangle _{\text{b}}\approx \frac{eq}{2\pi ^{2}a^{2}}\frac{%
\text{sgn}(\alpha _{0})(r/2a)^{2q_{\alpha }-1}}{\Gamma ^{2}(q_{\alpha }+1/2)}%
\int_{\mu }^{\infty }dz\,\frac{z^{2q_{\alpha }}}{\sqrt{z^{2}-\mu ^{2}}}
\notag \\
&&\quad \times \frac{\mu W_{q_{\alpha }-1/2,q_{\alpha
}+1/2}^{(+)}(z)-(z^{2}-\mu ^{2})/z}{z[I_{q_{\alpha
}-1/2}^{2}(z)+I_{q_{\alpha }+1/2}^{2}(z)]+2\mu I_{q_{\alpha
}-1/2}(z)I_{q_{\alpha }+1/2}(z)},  \label{j0intApex} \\
&&\langle j^{2}(x)\rangle _{\text{b}}\approx -\frac{eq}{\pi ^{2}a^{3}}\frac{%
\text{sgn}(\alpha _{0})(r/2a)^{2q_{\alpha }-1}}{(2q_{\alpha }+1)\Gamma
^{2}(q_{\alpha }+1/2)}\int_{\mu }^{\infty }dz\,\frac{z^{2q_{\alpha }+2}}{%
\sqrt{z^{2}-\mu ^{2}}}  \notag \\
&&\quad \times \frac{W_{q_{\alpha }-1/2,q_{\alpha }+1/2}^{(+)}(z)}{%
z[I_{q_{\alpha }-1/2}^{2}(z)+I_{q_{\alpha }+1/2}^{2}(z)]+2\mu I_{q_{\alpha
}-1/2}(z)I_{q_{\alpha }+1/2}(z)}.  \label{j2intApex}
\end{eqnarray}%
For a massless field these expressions reduce to Eq. (\ref{j02LargeRadm0}).
From here it follows that in the limit $r\rightarrow 0$ the boundary-induced
part vanishes when $|\alpha _{0}|<1/2-1/(2q)$ and diverges for $|\alpha
_{0}|>1/2-1/(2q)$. Notice that in the former case the irregular mode is
absent \ and the divergence in the latter case comes from the irregular
mode. This divergence is integrable. In the case $|\alpha _{0}|=1/2-1/(2q)$,
corresponding to $q_{\alpha }=1/2$, the boundary-induced VEV tends to a
finite limiting value. In particular, for the magnetic vortex in the
background Minkowski spacetime, the boundary-induced contribution diverges
as $r^{-2|\alpha _{0}|}$. In Fig. \ref{fig5}, the VEVs of the charge density
(left panel) and azimuthal current (right panel) are plotted for a massless
fermionic field inside a circular boundary as functions of the magnetic
flux. The graphs are plotted for $r/a=0.5$ and for several values of the
opening angle for the conical space. For a massless field the boundary-free
part in the VEV\ of the charge density vanishes and the charge density on
the left plot is induced by the boundary.

\section{Summary and conclusions}

\label{sec:Conc}

In this paper we have investigated the VEV of the fermionic current induced
by a magnetic flux string in a (2+1)-dimensional conical spacetime with a
circular boundary. The case of massive fermionic field is considered with
MIT bag boundary condition on the circle. In (2+1)-dimensional spacetime
there are two inequivalent irreducible representations of the Dirac
matrices. We have used the representation (\ref{DirMat}). The corresponding
results for the second representation are obtained by changing $m\rightarrow
-m$. Under this change, the boundary-free contribution to the VEV of the
azimuthal current is not changed, whereas the VEV of the charge density
changes the sign. For the evaluation of the expectation values we have
employed the direct mode summation method. The corresponding positive and
negative energy eigenspinors in the region outside the circular boundary are
given by Eqs. (\ref{psisig+}) and (\ref{psisig-}).

The VEV of the fermionic current in the boundary-free conical space is
investigated in Sect. \ref{sec:BoundFree}. For this geometry, under the
condition (\ref{condalf0}), there are no square integrable irregular modes.
In the case $|\alpha _{0}|>(1-1/q)/2$, the theory of von Neumann deficiency
indices leads to a one-parameter family of allowed boundary conditions at
the origin. Here we consider a special boundary condition that arises when
imposing bag boundary condition at a finite radius, which is then shrunk to
zero. The VEVs of the fermionic current for other boundary conditions on the
cone apex are evaluated in a similar way. The contribution of the regular
modes is the same for all boundary conditions and the formulae differ by the
parts related to the irregular modes. For the boundary condition under
consideration, the eigenspinors for the boundary-free geometry are obtained
from the corresponding functions in the region outside a circular boundary
with radius $a$, taking the limit $a\rightarrow 0$. They are presented by
Eq. (\ref{psi0}). When the magnetic flux, measured in units of the flux
quantum, is a half-integer, there is a special mode corresponding to the
angular momentum $j=-\alpha $, with the negative-energy eigenspinor given by
Eq. (\ref{psibetSp}).

In the boundary-free geometry, the regularized expressions of the VEVs are
given by Eqs. (\ref{j00reg1}) and (\ref{j02reg1}) for the charge density and
the azimuthal current, respectively, and the VEV of the radial component
vanishes. These VEVs are periodic functions of the parameter $\alpha $ with
the period equal to 1. So, if we present this parameter as (\ref{alf0}),
with $n_{0}$ being an integer number, then the VEVs are functions of $\alpha
_{0}$ alone. These functions are odd with respect to the reflection $\alpha
_{0}\rightarrow -\alpha _{0}$. Both charge density and azimuthal current
exhibit the jump structure at half-integer values for the ratio of the
magnetic flux to the flux quantum. Simple expressions for the renormalized
VEVs, Eqs. (\ref{j00Spren}) and (\ref{j02Spren}), are obtained in the
special case where the parameter $q$ related to the planar angle deficit is
an integer and the magnetic flux takes special values given by Eq. (\ref%
{alphaSpecial}). In the general case of parameters $\alpha $ and $q$, we
have derived two different representations for the renormalized VEVs. The
first one is based on Eq. (\ref{seriesI3}) and the corresponding expressions
for the charge density and azimuthal current have the forms (\ref{j02ren2})
and (\ref{j00ren2}). The second representation is obtained by using the
Abel-Plana summation formula and the corresponding expressions are given by
Eqs. (\ref{j00renb}), (\ref{j02renb}). For a massless field the VEV of the
charge density vanishes for points outside the magnetic vortex and the VEV
of the azimuthal current behaves as $r^{-3}$. For a massive field, for
points near the vortex the VEVs behave as $1/r$ and $1/r^{3}$ for the charge
density and the azimuthal current, respectively. At distances larger than
the fermion Compton wavelength, the VEVs decay exponentially with the decay
rate depending on the opening angle of the cone. The total charge induced by
the magnetic vortex does not depend on the angle deficit of the conical
space and is given by Eq. (\ref{Q1}). In the special case of a magnetic flux
in Minkowski spacetime, the formulae for the VEVs of the charge density and
the azimuthal current reduce to Eqs. (\ref{FCq1}), or equivalently, Eqs. (%
\ref{j02renbm0}). In Appendix \ref{sec:App2New} we show the equivalence of
Eqs. (\ref{j02renbm0}) to the expressions for the VEVs of charge density and
azimuthal current known from the literature.

The effects of a circular boundary on the VEV of the fermionic current are
considered in Sect. \ref{sec:ExtFC} for the exterior region. From the point
of view of the physics in this region, the circular boundary can be
considered as a simple model for the defect core. The mode sums of the
charge density and the azimuthal current are given by Eqs. (\ref{j02Ext})
and the radial component vanishes. In order to extract from the VEVs the
contributions induced by the boundary, we have subtracted the boundary-free
parts. Rotating the integration contours in the complex plane, we have
derived rapidly convergent integral representations for the boundary-induced
contributions, Eqs. (\ref{j002b}) and (\ref{j21}). These formulae are
simplified in the case of a massless field with expressions (\ref{J02Extm0}%
). In the exterior region, the total VEV of the fermionic current is a
continuous function of the magnetic flux. Note that unlike the boundary-free
part, the boundary-induced part in the VEV\ of the charge density for a
massless field is not zero. The parts in the VEVs induced by the boundary
are periodic functions of the magnetic flux with the period equal to the
flux quantum. These parts vanish for the special case of the magnetic flux
corresponding to $|\alpha _{0}|=1/2$. In the limit when the radius of the
circle goes to zero and for $|\alpha _{0}|<1/2$, for a massive field the
boundary-induced contributions in the exterior region behave as $%
a^{2q_{\alpha }}$, with $q_{\alpha }$ defined by Eq. (\ref{qalfa}). For a
massless field the corresponding asymptotics are given by Eqs. (\ref%
{j0bExta0m0})-(\ref{j2bExtm0b}). At large distances from the boundary and
for a massive field, the contributions coming from the boundary decay
exponentially [see Eqs. (\ref{j02LargeDist})]. In the same limit and for a
massless field, the boundary-induced VEV in the charge density decays as $%
(a/r)^{2q_{\alpha }}$. For the azimuthal current, the contribution induced
by a circular boundary behaves as $(a/r)^{2q_{\alpha }+1}$ when $q_{\alpha
}>1/2$, and like $(a/r)^{4q_{\alpha }}$ for $q_{\alpha }<1/2$. Note that,
when the circular boundary is present, the VEVs of physical quantities in
the exterior region are uniquely defined by the boundary conditions and by
the bulk geometry. This means that if we consider a non-trivial core model
for both conical space and magnetic flux with finite thickness $b<a$ and
with the line element (\ref{ds21}) in the region $r>b$, the results in the
region outside the circular boundary will not be changed.

The boundary-induced VEVs in the region inside a circular boundary are
studied in Sect. \ref{sec:Int}. The corresponding mode sums for the charge
density and the azimuthal current are given by Eq. (\ref{j2Int}). They
contain series over the zeros of the function given by Eq. (\ref{gamVal}).
For the summation of these series we have employed a variant of the
generalized Abel-Plana formula. The latter allowed us to extract explicitly
from the VEVs the parts corresponding to the conical space without
boundaries and to present the contributions induced by the circle in terms
of exponentially convergent integrals. In the interior region the
boundary-induced parts in the renormalized VEVs of the charge density and
the azimuthal current are given by Eqs. (\ref{j0int1}) and (\ref{j2int1}).
For a massless fermionic field these formulae are reduced to Eqs. (\ref%
{j02intm0}). For large values of the circle radius and for a massive field
the boundary-induced contribution decay exponentially [Eqs. (\ref%
{j02LargeRad})]. In the case of a massless field, the VEVs decay as $%
(r/a)^{2q_{\alpha }-1}$, for both charge density and azimuthal current. For
points near the apex of the cone, the leading terms in the corresponding
asymptotic expansions are given by Eqs. (\ref{j0intApex}) and (\ref%
{j2intApex}). In particular, the boundary-induced parts vanish at the apex
when $|\alpha _{0}|<1/2-1/(2q)$ and diverge for $|\alpha _{0}|>1/2-1/(2q)$.

The formulas for the VEV of the fermionic current are easily generalized for
the case of a spinor field with quasiperiodic boundary condition (\ref{PerBC}%
) along the azimuthal direction. This problem is reduced to the one we have
considered by a gauge transformation. The corresponding expressions for the
VEVs are obtained from those given above changing the definition of the
parameter $\alpha $ by Eq. (\ref{Replace}).

The results obtained in the present paper can be applied for the evaluation
of the VEV of the fermionic current in graphitic cones. Graphitic cones are
obtained from graphene sheet if one or more sectors are excised. The opening
angle of the cone is related to the number of sectors removed, $N_{c}$, by
the formula: $\phi _{0}=2\pi (1-N_{c}/6)$ , with $N_{c}=1,2,\ldots ,5$ (for
the electronic properties of graphitic cones see, e.g., \cite{Lamm00} and
references therein). All these angles have been observed in experiments \cite%
{Kris97}. The electronic band structure of graphene close to the Dirac
points shows a conical dispersion $E(\mathbf{k})=v_{F}|\mathbf{k}|$, where $%
\mathbf{k}$ is the momentum measured relatively to the Dirac points and $%
v_{F}\approx 10^{8}$ cm/s represents the Fermi velocity which plays the role
of a speed of light. Consequently, the long-wavelength description of the
electronic states in graphene can be formulated in terms of the Dirac-like
theory in (2+1)-dimensional spacetime. The corresponding excitations are
described by a pair of two-component spinors, corresponding to the two
different triangular sublattices of the honeycomb lattice of graphene (see,
for instance, \cite{Cast09}). In both cases of finite and truncated
graphitic cones the corresponding 2-dimensional surface has a circular
boundary. As the Dirac field lives on the cone surface, it is natural to
impose bag boundary condition (\ref{BCMIT}) on the bounding circle which
ensures the zero fermion flux through the edge of the cone. A more detailed
investigation of the fermionic current in graphitic cones, based on the
results of the present paper, will be presented elsewhere.

\section*{Acknowledgments}

E.R.B.M. and V.B.B. thank Conselho Nacional de Desenvolvimento Cient\'{\i}%
fico e Tecnol\'{o}gico (CNPq) and FAPES-ES/CNPq (PRONEX) for partial
financial support. A.A.S. was supported by Conselho Nacional de
Desenvolvimento Cient\'{\i}fico e Tecnol\'{o}gico (CNPq) and by the Armenian
Ministry of Education and Science Grant No. 119.

\appendix

\section{Integral representations}

\label{sec:IntRep}

In this section we derive two integral representations for the function $%
\mathcal{I}(q,\alpha _{0},z)$ defined by Eq. (\ref{seriesI0}). In the first
approach, we use the integral representation for the modified Bessel
function $I_{\beta _{j}}(z)$ (see formula 9.6.20 in Ref. \cite{hand}). To be
allowed to replace the order of the integration and the summation over $j$,
we integrate by parts the first term in this representation:%
\begin{equation}
I_{\beta _{j}}(z)=\frac{\sin (\pi \beta _{j})}{\pi \beta _{j}}e^{-x}+\frac{z%
}{\pi }\int_{0}^{\pi }dy\sin y\frac{\sin (\beta _{j}y)}{\beta _{j}}e^{z\cos
y}-\frac{\sin (\pi \beta _{j})}{\pi }\int_{0}^{\infty }dye^{-z\cosh y-\beta
_{j}y}.  \label{intRepI}
\end{equation}%
Substituting (\ref{intRepI}) into Eq. (\ref{seriesI0}) and interchanging the
order of the summation and integration, we apply the formula \cite{Prud86}%
\begin{equation}
\sum_{j}\frac{\sin (\beta _{j}y)}{\beta _{j}}=(-1)^{l}\frac{\pi \cos \left[
(2l+1)\pi (\alpha _{0}-1/2q)\right] }{q\cos [\pi (\alpha _{0}-1/2q)]},
\label{sinnu}
\end{equation}%
for $2l\pi /q<y<(2l+2)\pi /q$. For the first term in the right-hand side of (%
\ref{intRepI}) one has $y=\pi $. For this term, when $q=2l$ with $%
l=1,2,\ldots $, the corresponding summation formula has the form%
\begin{equation}
\sum_{j}\frac{\sin (\pi \beta _{j})}{\beta _{j}}=(-1)^{q/2}\frac{\pi }{q}%
\cos (q\pi \alpha _{0})\tan [\pi (\alpha _{0}-1/2q)].  \label{sinnu2}
\end{equation}%
Finally, for the series corresponding to the last term in Eq. (\ref{intRepI}%
) we have%
\begin{equation}
\sum_{j}\sin (\pi \beta _{j})e^{-\beta _{j}y}=\frac{f(q,\alpha _{0},y)}{%
\cosh (qy)-\cos (q\pi )},  \label{sinnu3}
\end{equation}%
with the notation%
\begin{eqnarray}
f(q,\alpha _{0},y) &=&\cos \left[ q\pi \left( 1/2-\alpha _{0}\right) \right]
\cosh \left[ \left( q\alpha _{0}+q/2-1/2\right) y\right]  \notag \\
&&-\cos \left[ q\pi \left( 1/2+\alpha _{0}\right) \right] \cosh \left[
\left( q\alpha _{0}-q/2-1/2\right) y\right] .  \label{fqualf}
\end{eqnarray}

Combining the formulae given above, we find the following integral
representation for the series (\ref{seriesI0}):%
\begin{eqnarray}
&& \mathcal{I}(q,\alpha _{0},z) =\frac{e^{z}}{q}-\frac{1}{\pi }%
\int_{0}^{\infty }dy\frac{e^{-z\cosh y}f(q,\alpha _{0},y)}{\cosh (qy)-\cos
(q\pi )}  \notag \\
&& \qquad +\frac{2}{q}\sum_{l=1}^{p}(-1)^{l}\cos [2\pi l(\alpha
_{0}-1/2q)]e^{z\cos (2\pi l/q)},  \label{seriesI3}
\end{eqnarray}%
with $2p<q<2p+2$. In the case $q=2p$, the term
\begin{equation}
-(-1)^{q/2}\frac{e^{-z}}{q}\sin (q\pi \alpha _{0}),  \label{replaced}
\end{equation}%
should be added to the right-hand side of Eq. (\ref{seriesI3}). Note that
for $1\leqslant q<2$, the last term on the right-hand side of Eq. (\ref%
{seriesI3}) is absent. Formula (\ref{seriesI3}) is simplified in the case $%
q=1$:%
\begin{equation}
\mathcal{I}(1,\alpha _{0},z)=e^{z}-\frac{\sin (\pi \alpha _{0})}{\pi }%
\int_{0}^{\infty }dye^{-z\cosh y}\frac{\cosh [\left( 1/2-\alpha _{0}\right)
y]}{\cosh (y/2)}.  \label{seriesIq1}
\end{equation}

In the special case (\ref{alphaSpecial}) with an integer $q$, the integral
term in Eq. (\ref{seriesI3}) vanishes. For $q=2p+1$ one finds $\mathcal{I}%
(q,\alpha ,z)=(2/q)\sum_{l=0}^{(q-1)/2\prime }e^{z\cos (2\pi l/q)}$, where,
as before, the prime on the summation sign mens that the term with $l=0$
should be halved. For even values of $q$, by taking into account the
additional term (\ref{replaced}), we find $\mathcal{I}(q,\alpha
,z)=(2/q)\sum_{l=0}^{q/2\prime }e^{z\cos (2\pi l/q)}-e^{-z}/q$. Note that
for an integer $q$, from definition (\ref{seriesI0}) one has $\mathcal{I}%
(q,\alpha ,z)=2\sum_{n=0}^{\infty \prime }I_{qn}(z)$. Now, we can see that
in the special case under consideration, formula (\ref{seriesI3}) coincides
with Eq. (\ref{SerSp}).

\begin{figure}[tbph]
\begin{center}
\epsfig{figure=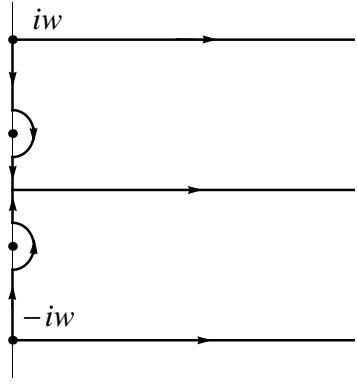,width=5.cm,height=5.cm}
\end{center}
\caption{Integration contour in the complex plane $z$ in the derivation of
summation formula~(\protect\ref{SumForm2}).}
\label{fig6}
\end{figure}

We can give an alternative integral representation of the series (\ref%
{seriesI0}) by using the Abel-Plana summation formula in the form (see \cite%
{Most97,Saha08Book})%
\begin{equation}
\sum_{n=0}^{\infty }f(n+1/2)=\int_{0}^{\infty }dz\,f(z)-i\int_{0}^{\infty
}dz\,\frac{f(iz)-f(-iz)}{e^{2\pi z}+1}.  \label{APF}
\end{equation}%
First of all we write this formula in the form more appropriate for the
application to Eq. (\ref{seriesI0}). Let us consider the series%
\begin{equation}
\sum_{j}f(|j+u|+v\epsilon _{j})=\sum_{\delta =\pm 1}\sum_{n=0}^{\infty
}f(n+1/2+\delta |u+v|),  \label{Ser1}
\end{equation}%
where $|u|\leqslant 1/2$, $|v|\leqslant 1/2$. Applying formula (\ref{APF}),
we get%
\begin{equation}
\sum_{j}f(|j+u|+v\epsilon _{j})=\sum_{\delta =\pm 1}\int_{0}^{\infty
}dz\,f(z+\delta w)-i\sum_{\delta =\pm 1}\int_{0}^{\infty }dz\,\frac{\delta
f(i\delta (z+iw))+\delta f(i\delta (z-iw)}{e^{2\pi z}+1},  \label{Ser2}
\end{equation}%
with the notation $w=|u+v|$. Introducing new integration variables, we
present this formula in the form
\begin{eqnarray}
\sum_{j}f(|j+u|+v\epsilon _{j}) &=&2\int_{0}^{\infty }dzf(z)-\int_{0}^{w}dz
\left[ f(z)-f(-z)\right]  \notag \\
&&-i\sum_{\delta =\pm 1}\int_{i\delta w}^{\infty +i\delta w}dz\,\frac{%
f(iz)-f(-iz)}{e^{2\pi (z-i\delta w)}+1}.  \label{SumForm1}
\end{eqnarray}%
Now, deforming the integration contour, we write the integral along the
half-line $(i\delta w,\infty +i\delta w)$ in the complex plane $z$ as the
sum of the integrals along the segment $(i\delta w,0)$ and along $(0,\infty
) $. At this step we note that in the case $1/2<w<1$ the integrand has a
pole at $z=i\delta (w-1/2)$. We exclude the poles by small semicircles in
the right-half plane with radius tending to zero (see Fig. \ref{fig6}). The
sum of the integrals along the segments $(iw,0)$ and $(-iw,0)$ cancels the
second integral in the right-hand side of Eq. (\ref{SumForm1}) and we get
the following result%
\begin{equation}
\sum_{j}f(|j+u|+v\epsilon _{j})=A+2\int_{0}^{\infty
}dxf(x)-i\int_{0}^{\infty }dy\,\sum_{\delta =\pm 1}\frac{f(iy)-f(-iy)}{%
e^{2\pi (y+i\delta |u+v|)}+1},  \label{SumForm2}
\end{equation}%
where $A=0$ for $0\leqslant |u+v|\leqslant 1/2$, and%
\begin{equation}
A=f(1/2-|u+v|)-f(|u+v|-1/2),  \label{A}
\end{equation}%
for $1/2<|u+v|<1$. The term $A$ comes from the contributions of the above
mentioned poles to the integrals.

We apply to the series in Eq. (\ref{seriesI0}) formula (\ref{SumForm2}) with
the function $f(z)=I_{qz}(x)$ and with the parameters $u=\alpha _{0}$, $%
v=-1/(2q)$. This leads to the following result%
\begin{eqnarray}
&&\mathcal{I}(q,\alpha _{0},x)=A(q,\alpha _{0},x)+\frac{2}{q}%
\int_{0}^{\infty }dz\,I_{z}(x)\,  \notag \\
&&\qquad -\frac{4}{\pi q}\int_{0}^{\infty }dz\,{\mathrm{Re}}\left[ \frac{%
\sinh (z\pi )K_{iz}(x)}{e^{2\pi (z+i|q\alpha _{0}-1/2|)/q}+1}\right] ,
\label{Rep2}
\end{eqnarray}%
with $A(q,\alpha _{0},x)=0$ for $|\alpha _{0}-1/2q|\leqslant 1/2$, and
\begin{equation}
A(q,\alpha _{0},x)=\frac{2}{\pi }\sin [\pi (|q\alpha
_{0}-1/2|-q/2)]K_{|q\alpha _{0}-1/2|-q/2}(x),  \label{Aq}
\end{equation}%
for $1/2<|\alpha _{0}-1/2q|<1$. Note that in Eq. (\ref{Rep2}) the function $%
K_{iz}(x)$ is real and the real part in the integrand can be written
explicitly by observing that%
\begin{equation}
{\mathrm{Re\,}}(e^{u+iv}+1)^{-1}=\frac{1}{2}\frac{e^{-u}+\cos v}{\cosh
u+\cos v}.  \label{Repart}
\end{equation}

In the special case $q=1$, from (\ref{Rep2}) we have%
\begin{eqnarray}
&&\mathcal{I}(1,\alpha _{0},x)=A(1,\alpha _{0},x)+2\int_{0}^{\infty
}dz\,I_{z}(x)\,  \notag \\
&&\qquad +\frac{4}{\pi }\int_{0}^{\infty }dz\,{\mathrm{Re}}\left[ \frac{%
\sinh (z\pi )K_{iz}(x)}{e^{2\pi (z-i\alpha _{0})}-1}\right] ,  \label{Repq2}
\end{eqnarray}%
with $A(1,\alpha _{0},x)=-(2/\pi )\sin (\pi \alpha _{0})K_{\alpha _{0}}(x)$
in the case $-1/2<\alpha _{0}<0$ and $A(1,\alpha _{0},x)=0$ for $0\leqslant
\alpha _{0}\leqslant 1/2$. Note that the last term in the right-hand side is
an even function of $\alpha _{0}$.

\section{Comparison with the previous results in the Minkowski bulk}

\label{sec:App2New}

The VEVs of the fermionic current induced by a magnetic flux in
(2+1)-dimensional Minkowski spacetime have been previously investigated in
Ref. \cite{Flek91}. The expressions for the VEVs of charge density and the
azimuthal current derived in this paper have the form (in our notations)%
\begin{eqnarray}
\langle j^{0}(x)\rangle _{0,\text{ren}} &=&-em\frac{\sin \left( \pi \alpha
_{0}\right) }{\pi ^{3}}\int_{m}^{\infty }dk\frac{kK_{\alpha _{0}}^{2}(kr)}{%
\sqrt{k^{2}-m^{2}}},  \notag \\
\langle j^{2}(x)\rangle _{0,\text{ren}} &=&e\frac{\sin (\pi \alpha _{0})}{%
\pi ^{3}}\int_{m}^{\infty }dk\,k^{3}\frac{K_{\alpha _{0}}^{2}(kr)-K_{\alpha
_{0}-1}(kr)K_{\alpha _{0}+1}(kr)}{\sqrt{k^{2}-m^{2}}}.  \label{j02Prev}
\end{eqnarray}%
In this section, we show that these representations are equivalent to Eqs. (%
\ref{j02renbm0}). First we consider the charge density. As the first step,
we use the integral representation for the square of the Macdonald function:%
\begin{equation}
K_{\alpha _{0}}^{2}(x)=\frac{1}{2}\int_{0}^{\infty }\frac{dz}{z}%
e^{-x^{2}/2z-z}K_{\alpha _{0}}(z).  \label{IntRepK2}
\end{equation}%
This formula is easily obtained from the integral representation of the
product of two Macdonald functions given in Ref. \cite{Wats44} (page 439).
Inserting Eq. (\ref{IntRepK2}) into (\ref{j02Prev}), after the change of the
order of integrations, the integral over $k$ is taken simply and for the
charge density we obtain the result given in Eq. (\ref{j02renbm0}).

In order to see the equivalence of the representations for the azimuthal
current, we present the expression with the Macdonald functions in the form%
\begin{equation}
K_{\alpha _{0}}^{2}(kr)-K_{\alpha _{0}-1}(kr)K_{\alpha _{0}+1}(kr)=-\frac{2}{%
r^{2}}\int_{r}^{\infty }dxxK_{\alpha _{0}}^{2}(kx).  \label{IntRepK3}
\end{equation}%
After substituting this into Eq. (\ref{j02Prev}), we use the integral
representation (\ref{IntRepK2}). Then, we first take the integral over $k$
and after that the integral over $x$. As a result, the integral
representation (\ref{j02renbm0}) for the azimuthal current is obtained.
Hence, we have shown that, in the special case of Minkowski bulk, our
results for the VEVs of the charge density and azimuthal current, given by
Eq. (\ref{j02renbm0}), agree with those from the literature. Note that we
have also derived alternative representations (\ref{FCq1}). For the
numerical evaluation the latter are more convenient.

\section{Contribution of the mode with $j=-\protect\alpha $}

\label{sec:App2}

When the parameter $\alpha $ is equal to a half-integer, the contribution of
the mode with $j=-\alpha $ to the VEV of the fermionic current should be
evaluated separately. Here we consider the region inside a circle with
radius $a$. Similar to Eq. (\ref{psibetSp}), the negative-energy eigenspinor
for this mode has the form%
\begin{equation}
\psi _{\gamma ,-\alpha }^{(-)}(x)=\frac{b_{0}}{\sqrt{r}}e^{iq\alpha \phi
+iEt}\left(
\begin{array}{c}
\frac{\gamma e^{-iq\phi /2}}{E+m}\sin (\gamma r-\gamma _{0}) \\
e^{iq\phi /2}\cos (\gamma r-\gamma _{0})%
\end{array}%
\right) ,  \label{psigamSp}
\end{equation}%
where $\gamma _{0}$ is defined after Eq. (\ref{psibetSp}). From boundary
condition (\ref{BCMIT}) it follows that the eigenvalues of $\gamma $ are
solutions of the equation%
\begin{equation}
m\sin (\gamma a)+\gamma \cos (\gamma a)=0.  \label{modeqSp}
\end{equation}%
The positive roots of this equation we denote by $\gamma _{l}=\gamma a$, $%
l=1,2,\ldots $. From the normalization condition, for the coefficient in (%
\ref{psigamSp}) one has%
\begin{equation}
b_{0}^{2}=\frac{E+m}{aE\phi _{0}}\left[ 1-\sin (2\gamma a)/(2\gamma a)\right]
^{-1}.  \label{b02}
\end{equation}

Using Eq. (\ref{psigamSp}), for the contributions of the mode under
consideration to the VEVs of the fermionic current we find:%
\begin{eqnarray}
\langle j^{0}(x)\rangle _{j=-\alpha } &=&\frac{e}{ar\phi _{0}}%
\sum_{l=1}^{\infty }\frac{1+\mu \left[ \gamma _{l}\sin (2\gamma _{l}r/a)-\mu
\cos (2\gamma _{l}r/a)\right] /(aE)^{2}}{1-\sin (2\gamma _{l})/(2\gamma _{l})%
},  \notag \\
\langle j^{2}(x)\rangle _{j=-\alpha } &=&-\frac{e}{ar^{2}\phi _{0}}%
\sum_{l=1}^{\infty }\frac{\gamma _{l}}{(aE)^{2}}\frac{\mu \sin (2\gamma
_{l}r/a)+\gamma _{l}\cos (2\gamma _{l}r/a)}{1-\sin (2\gamma _{l})/(2\gamma
_{l})},  \label{j02SpAp}
\end{eqnarray}%
where $(aE)^{2}=\gamma _{l}^{2}+\mu ^{2}$ and $\mu =ma$. The part
corresponding to the radial component vanishes. We assume the presence of a
cutoff function. For the summation of the series in Eqs. (\ref{j02SpAp}), we
use the Abel-Plana-type formula%
\begin{equation}
\sum_{l=1}^{\infty }\frac{\pi f(\gamma _{l})}{1-\sin (2\gamma _{l})/(2\gamma
_{l})}=-\frac{\pi f(0)/2}{1/\mu +1}+\int_{0}^{\infty
}dz\,f(z)-i\int_{0}^{\infty }dz\frac{f(iz)-f(-iz)}{\frac{z+\mu }{z-\mu }%
e^{2z}+1}.  \label{SumFormAp}
\end{equation}%
Eq. (\ref{SumFormAp}) is obtained from more general summation formula given
in \cite{Rome02} (see also \cite{Saha08Book}) taking $b_{1}=0$ and $%
b_{2}=-1/\mu $. For the functions $f(z)$ corresponding to Eq. (\ref{j02SpAp}%
) one has $f(0)=0$. The contribution of the last term in Eq. (\ref{SumFormAp}%
) to $\langle j^{\nu }(x)\rangle _{j=-\alpha }$ is finite in the limit when
the cutoff is removed. Noting that for both series in Eq. (\ref{j02SpAp})
the function $f(z)$ is an even function, we conclude that the last term in (%
\ref{SumFormAp}) does not contribute to the both charge density and
azimuthal current. For the remaining parts coming from the second term in
the right-hand side of Eq. (\ref{SumFormAp}) one has%
\begin{eqnarray}
\langle j^{0}(x)\rangle _{j=-\alpha } &=&\frac{e}{\pi r\phi _{0}}%
\int_{0}^{\infty }d\gamma \left[ 1+h(\gamma )\right] ,  \notag \\
\langle j^{2}(x)\rangle _{j=-\alpha } &=&-\frac{e}{\pi r^{2}\phi _{0}}%
\int_{0}^{\infty }d\gamma \left[ \cos (2\gamma r)+h(\gamma )\right] ,
\label{j02Ap2}
\end{eqnarray}%
with $h(\gamma )=mE^{-2}\left[ \gamma \sin (2\gamma r)-m\cos (2\gamma r)%
\right] $ and $E^{2}=\gamma ^{2}+m^{2}$. These parts do not depend on the
circle radius. Now, we can see that the integral with the function $h(\gamma
)$ is zero. The remained term in the charge density is subtracted by the
renormalization and the remaining integral in the expression of azimuthal
current is zero for $r>0$. Hence, we conclude that the term with $j=-\alpha $
does not contribute to the renormalized VEV of the fermionic current.

\end{document}